\documentclass[sigconf,authorversion]{acmart}

\AtBeginDocument{\providecommand\BibTeX{{\normalfont B\kern-0.5em{\scshape i\kern-0.25em b}\kern-0.8em\TeX}}}

\setcopyright{none}
\copyrightyear{2026}
\acmYear{2026}
\acmISBN{978-1-4503-XXXX-X/2018/06}

\usepackage{macros}
\usepackage{submission}

\usepackage[T1]{fontenc}
\usepackage{newtxtext}
\usepackage{newtxmath}
\usepackage[all]{foreign}
\usepackage{subcaption}
\usepackage[normalem]{ulem}
\usepackage{array}

\setmode{final}

\renewcommand{\etc}{\textit{etc.}\xspace}
\renewcommand{\ie}{\textit{i.e.}\xspace}
\renewcommand{\eg}{\textit{e.g.}\xspace}

\acmConference[Preprint]{}{HAL}{2026}

\author{Ahmed Ben Akouche}
\email{benakouche@isir.upmc.fr}
\orcid{0009-0006-7953-9368}
\affiliation{%
  \institution{Sorbonne Université, CNRS, Inserm, Institut des Systèmes Intelligents et de Robotique, ISIR}
  \city{F-75005 Paris}
  \country{France}
}

\author{G\'ery Casiez}
\email{gery.casiez@univ-lille.fr}
\orcid{0000-0003-1905-815X}
\affiliation{%
  \institution{Univ. Lille, CNRS, Inria, Centrale Lille, UMR 9189 CRIStAL}
  \postcode{F-59000}
  \city{Lille}
  \country{France}}

\author{Mathieu Nancel}
\email{mathieu.nancel@inria.fr}
\orcid{0000-0002-0804-728X}
\affiliation{%
  \institution{Univ. Lille, Inria, CNRS, Centrale Lille, UMR 9189 CRIStAL}
  \postcode{F-59000}
  \city{Lille}
  \country{France}
}

\author{Julien Gori}
\email{julien.gori@sorbonne-universite.fr}
\orcid{0000-0002-3272-9176}
\affiliation{%
  \institution{Sorbonne Université, CNRS, Inserm, Institut des Systèmes Intelligents et de Robotique, ISIR}
  \city{F-75005 Paris}
  \country{France}
}

\begin{document}

\title{\technique: Detecting Widgets from Pixels\\ on Desktop Interfaces}

\begin{abstract}
"Target-aware" pointing techniques, like Bubble Cursor or Semantic Pointing, outperform traditional pointing by leveraging knowledge of target locations. Yet the lack of application-agnostic widget geometry information limits their adoption across the desktop. We present TargetFinder, a computer vision-based system for real-time detection of GUI widgets. TargetFinder leverages several fine-tuned YOLO networks trained on a new dataset of 520 annotated desktop screenshots ($\sim$38,000 annotations) spanning Windows, macOS, Ubuntu, and web interfaces. TargetFinder uses lightweight screen monitoring and low-latency detection, achieving millisecond responsiveness suitable for interactive use. Evaluations show that TargetFinder outperforms the baseline methods (OmniParser and REMAUI), while system-wide implementations of Bubble Cursor and Semantic Pointing demonstrate the feasibility of deploying universal target-aware techniques that work across applications. We release the dataset, models, annotation tool, and an open-source library for research and applications.
\end{abstract}

%%
%% The code below is generated by the tool at http://dl.acm.org/ccs.cfm.
%% Please copy and paste the code instead of the example below.
%%



\maketitle

\section{Introduction}

Widgets, such as buttons, menus, sliders, or text fields, are interactive elements that compose Graphical User Interfaces (GUI). Their nature, position, and dimensional properties often remain encapsulated within application logic, and mostly inaccessible to other applications.
This limitation restricts the development of advanced interaction techniques across multiple domains~\cite{jansen2011}.
A prominent example is target-aware selection facilitation, \ie techniques requiring knowledge of all target positions and dimensions on a display, of which many have been proposed in the HCI literature.
This includes area cursors~\cite{grossman2005,findlater2010, kabbash1995, worden1997, chapuis2009, su2014, hertzum2007}, cursors that dynamically manipulate the control-display ratio~\cite{blanch2004,guiard2004}, expanding/movable targets~\cite{zhai2003, mcguffin2002, baudisch2003}, and techniques based on predicting user intent~\cite{asano2005, liu2017, ziebart2012}.
These techniques have been demonstrated only in application-specific prototypes where researchers maintain direct control over the interface, or within web browsers whose layout engine APIs readily expose target geometry~\cite{gajos2012}. The lack of accessible widget geometry prevents system-wide implementation of target-aware pointing techniques, limiting users to ``regular'' target-agnostic interactions~\cite{casiez2011}, despite research consistently demonstrating the superior performance of target-aware techniques~\cite{grossman2005,blanch2004,guiard2004}.
To our knowledge, the latter have seldom been evaluated in the wild, with the exception of a Bubble Cursor web-browser implementation cited in~\cite[ref. 8]{jansen2011} but no longer accessible.
Target-agnostic methods may also benefit from widget information; for example, the transfer function tuning in AutoGain~\cite{lee2020} could be made more precise with target size data.
Accessibility APIs should be the tool of choice for system-wide target-aware techniques, as they theoretically provide the necessary widget information, but are both OS and application-dependent\footnote{each OS offers distinct formats and capabilities, the granularity of exposed information is application-dependent making generic inspections unreliable, and DPI scaling inconsistencies can cause mismatches between reported and actual widget locations}.
Most critically, many third-party applications fail to implement accessibility support~\cite{patel2020,bi2022, MSActiveAccessibility, MSUIAutomation}, making reliable widget layout information difficult to obtain.
For example, sometimes a widget's higher-level container or the window itself is returned~\cite{malacria13}.

We present \technique, a cross-platform library built around a computer vision-based tool for identifying interface widgets via real-time analysis of desktop environments.
We fine-tune object detectors from the YOLO family~\cite{YOLO} on a novel dataset of annotated desktop interface screenshots spanning multiple OS's and application types.

\technique continuously monitors the screen. Significant changes are detected using fast comparison of low-resolution screenshots, upon which a full-resolution screenshot is processed by the YOLO model, which returns within milliseconds the bounding box coordinates of all visible interface widgets, along with their types: buttons, text fields, sliders \etc
The reasonably short processing latency (less than 210\,ms on a consumer-grade laptop;
see ~\autoref{sub:limit:realtime}) makes the system responsive enough for most interactive applications without introducing perceptible delays in user workflows, and realistically enables real-time use.

A fundamental advantage of our pure computer vision-based approach is its independence from application internals, accessibility APIs, or platform-specific frameworks.
Thanks to the diversity of the training set, \technique works across operating systems (Windows, macOS, Ubuntu), application frameworks (native, web, third-party), and even legacy software. The contributions of this article are as follows:
\begin{description}
\item[New annotated widget dataset:] We present a new dataset of 520 annotated screenshots covering Windows, macOS, Ubuntu, as well as web interfaces (130 images per category, each image containing multiple widgets).
The entire dataset consists of about 38,000 annotations.
We additionally provide a test set of 71 annotated screenshots, with 4,206 annotations, from older software covering diverse visual themes (light, dark, \etc) and varying image resolutions.
\item[Trained YOLO networks and evaluation:] We trained several YOLO models on our dataset and evaluated the performance of each of them to optimize \technique. We then present a comprehensive comparison of multiple widget detection methods.
\technique outperforms the existing baselines (OmniParser and REMAUI) evaluated in our study.
We also analyze generalization, including cross-OS transfer, desktop/mobile adaptability, and transfer to legacy software.
\item[Library:] We provide a library that includes a Python module (\texttt{TargetFinder}) for real-time widget detection, enabling easy integration into external applications or interaction techniques.
The library is cross-platform and hosted on PyPI, provides a command-line interface to visualize detections in real time through an overlay, and includes the two interaction techniques, Bubble Cursor and Semantic Pointing, tested on several versions of Windows and Ubuntu with X11.
\item[System-wide proof-of-concept:] We implemented system-wide versions of two target-aware selection techniques, the Bubble Cursor~\cite{grossman2005} and Semantic Pointing~\cite{blanch2004}, installable on any Windows 10-11 and Ubuntu-X11 machines and demonstrate the real-time applicability of \technique.
\end{description}

All the above contributions are released as open resources, including the public datasets\footnote{\url{https://osf.io/fr6y4/overview}} and the library\footnote{\url{https://github.com/AHMEDBENAKOUCHE/target_finder_toolkit.git}}.
This work highlights the potential of computer vision to overcome traditional barriers to accessing GUI widgets programmatically, and opens new perspectives for developing interaction techniques that do not require accessing or modifying application source code.

\section{Related Work}
\subsection{Goals of GUI widget detection}
Most research on GUI widget detection has focused on mobile contexts. The primary goal has been interface reconstruction \ie the automatic generation of functional code from screenshots or prototypes. For instance, REMAUI leverages widget detection to reverse engineer mobile user interfaces, converting images into working UI code~\cite{nguyen2015}. \citeauthor{daniel2020} review a variety of systems developed for automatic GUI code generation from images~\cite{daniel2020}, including work such as Pix2Code~\cite{tony2018}. GUI widget detection is also used in other downstream tasks; for example, predicting the tappability of widgets~\cite{swearngin2019}, or content description generation~\cite{bai2021}.

\medskip

In the desktop context, the main goal is to gather widget information for artificial agents that interact directly with GUIs \eg OS-ATLAS~\cite{wu2024}, WinClick~\cite{zheng2025}, see also~\cite{matti2023, martinez2024}.
These approaches do not aim for exhaustive detection of GUI widgets. Instead, they target instruction-driven interaction tasks where the agent must locate one or several specific widgets relevant to the requested action. An exception is OmniParser~\cite{lu2024omniparser}, which performs exhaustive widget detection and associates user actions to them.

\subsection{Identifying widgets from pixels via computer vision}
Identifying GUI widgets from images has been addressed through classical image processing and deep learning approaches. Classical image processing approaches rely on low-level operations like edge detection, dilation, and component segmentation. They are computationally efficient and unsupervised, \ie they do not require annotated training data, but often involve manual tuning of hyperparameters, and can lack robustness.
These techniques tend to over-segment small or decorative non-interactive elements, such as tiny icons or text, which can clutter the results and reduce practical usefulness. An example is REMAUI~\cite{nguyen2015}, which detects non-textual elements using the Canny edge detection~\cite{canny1986}, followed by dilation to merge close components and close gaps in contours.

Object detection models based on supervised neural networks, such as YOLO~\cite{altinbas2022}, Faster R-CNN~\cite{chen2020}, and SSD~\cite{zhang2021screen}, offer greater flexibility and accuracy.
Their main strength is their ability to detect widgets without having to rely on rules or intuitions specified by researchers.
Many of these models are pre-trained on large datasets, so only fine-tuning on GUI-specific data is needed. The main limit of these supervised methods is that they require labeled data \ie datasets of images where all GUI widgets are identified. Building or acquiring such datasets is costly, since a single desktop GUI image may contain more than 200 widgets.
While several neural network-based object detection for exhaustive widget identification exist for mobile interface recognition~\cite{xie2020,altinbas2022,chen2020}, these trained models fail to generalize to desktop interfaces\footnote{We tested the model from \citeauthor{chen2020}~\cite{chen2020}, trained on 40k images of the RICO dataset, with poor results; examples are provided in the supplementary materials.}.
In the desktop domain, OmniParser~\cite{lu2024omniparser} uses a YOLOv8 model trained on automatically labeled web screenshots (derived from the DOM), combined with optical character recognition (OCR) and icon features extracted with BLIP-v2 to support a vision-enabled language model (GPT-4V).
Another notable work is that of \citeauthor{martinez2024}~\cite{martinez2024}, which employs a multi-model detection pipeline consisting of five separately trained YOLOv8 networks, each specialized in detecting widgets at a different hierarchical level. We include both systems as baselines in \autoref{sub:comparison_baselines}.
Hybrid approaches, that combine classical computer vision techniques with deep learning, like UIED~\cite{xie2020}, also exist.

Segmentation models such as SAM~\cite{ravi2024} and MobileSAM~\cite{zhang2023} are an alternative solution. These models generate precise masks for visible objects, which can help refine the boundaries of GUI widgets. However, limitations are that they segment everything visible in the image, including non-interactive regions and are computationally demanding. A final work that should be mentioned, and that predates modern computer vision, is Prefab~\cite{dixon2010}, which attempts to recover widget structure directly from pixels from manually constructed visual prototypes. Prefab was namely used for a  Windows 7 implementation of a Bubble Cursor~\cite{dixon2012}.

\subsection{Alternate methods to detect widgets\label{sub:rw:alternate_methods}}
Other than image processing, we found three other methods in the literature to detect widget information. In all three cases, the author's goal was to perform a field study of Fitts' law, which requires information about the target sizes and locations.\\
Several works have exploited the \textbf{Accessibility/Automation API}\footnote{On Windows, Microsoft Active Accessibility (MSAA) has been originally introduced to expose information about UI elements~\cite{MSActiveAccessibility}. From 2005, it has been progressively replaced by UI Automation (UIA) but both remain available today~\cite{MSUIAutomation}. They are both natively supported by languages such as C++ and C\# but other languages, such as Java and Node.js, provide limited or complex support. Frameworks, like Qt, do not natively support UIA. In all cases, accessibility is not enabled by default and the developer needs to write more or less code to support accessibility in an application. }. Chapuis and colleagues \cite{chapuis2007,chapuis2005} tried to access target size and location via the accessibility API on Mac OS X, and Linux with GTK, but reported having to use undocumented functionalities to access additional information such as stacking order, and report being able to extract about 22\% of possible target bounds. \citeauthor{hurst2010} report that the Windows accessibility API alone enabled them to find 74\% of on-screen targets~\cite{hurst2010}, while \citeauthor{evans2012} achieved 67\% precision in their work with the Windows automation API~\cite{evans2012}. In all these cases, the accessibility API was always leveraged to determine the selected widget, and not all widgets on screen.\\
\citeauthor{evans2012} used the automation API to prelabel selected widgets before submitting them to \textbf{Amazon mechanical Turk} for final annotation~\cite{evans2012}. To improve reliability, each image was labeled by three independent workers, though the agreement rate was according to the authors unexpectedly low, at around 40\%.\\
\textbf{Application-specific implementations}: \citeauthor{gajos2012} developed a Firefox extension that leveraged the web browser's layout engine to access webpage geometry, including most clickable elements~\cite{gajos2012}. While they did not report the accuracy of their method, it is likely to be high\footnote{We developed a similar browser extension and obtained encouraging results, though we do not report on it here, as it does not offer a system-wide solution.}. Before that, \citeauthor{hurst2007} had developed a similar system for the open-source image editing software GIMP~\cite{hurst2007}.

\subsection{Existing GUI datasets with labeled elements}
The majority of existing GUI datasets focus on mobile user interfaces, see for example {RICO}~\cite{deka2017} and {VINS}~\cite{bunian2021}.
Fully annotated public desktop GUI datasets are scarce. To date, only a few efforts have attempted to provide comprehensive annotations:
\begin{itemize}
\item The \textbf{Waltteri Github} repository~\cite{waltteri2020} includes 51 fully annotated desktop screenshots. While annotation documentation is limited, the labels appear to be precise and complete.
\item The dataset proposed by \textbf{\citeauthor{martinez2024}}~\cite{martinez2024} contains 100 manually annotated desktop UI images.\item \textbf{DAWOD}~\cite{zhu2023} is a dataset of synthetic GUI images, but is not publicly available.
\item OmniParser~\cite{lu2024omniparser} relies on a large collection of automatically labeled web screenshots (67k, derived from the DOM). However, the dataset is not publicly available, covers only web interfaces, and does not contain screenshots of native desktop GUIs.
\end{itemize}

Other datasets, such as \textbf{OmniACT}~\cite{kapoor2024} and \textbf{OS-ATLAS}~\cite{wu2024}, provide partially annotated desktop and web interface images, primarily aimed at training autonomous agents, but are not designed for comprehensive GUI element detection.
OS-ATLAS leverages accessibility APIs for desktop and mobile platforms, and HTML parsing for web interfaces, but images are intentionally partially labeled to preserve diversity.
For OmniACT, desktop annotations are manually created by human annotators using a dedicated PyQt interface without aiming for full coverage.
The \textbf{UI-Vision} benchmark~\cite{nayak2025} was designed for GUI perception and interaction in desktop environments. The associated paper claims high-quality annotations including bounding boxes, UI element labels, and actions such as clicks and drags, but the publicly available data includes only a few annotations per image, which are cropped to application windows.

\section{Dataset acquisition}
\subsection{Tool\label{subs:tool}}
We developed an annotation tool called \wat \footnote{\url{https://github.com/AHMEDBENAKOUCHE/ui-widget-annotator-a-tool-for-fast-and-efficient-gui-annotation.git}} for semi-automatic annotations of GUI images.
The user can use several prelabeling methods\footnote{These include REMAUI and UIED, MobileSAM, and now, our trained YOLO model.} and interactively adjust, correct, or complete the annotations directly on the image using direct manipulation (see \autoref{fig:annotation_tool}).
In addition, \wat supports multi-class pre-classification through a predefined list of widget categories (buttons, sliders, text fields, \etc), using three existing convolutional neural networks (CNNs) trained on mobile data from~\cite{tezansahu2020}.
We estimate that the tool reduced annotation time by a factor two.

\subsection{Procedure}
We sourced partially annotated desktop images from OS-ATLAS~\cite{wu2024} and OmniACT~\cite{kapoor2024}.
We used OS-ATLAS for images of native desktop GUIs, which contains data from Windows, macOS, and Ubuntu operating systems for various versions of the platforms. All images were annotated from scratch using \wat.
The web category was populated with images from OmniACT.
The partial web annotations were generally accurate and visually aligned, so we retained the existing bounding boxes and completed the labeling.

The annotation process involved the initial pre-labeling phase provided by \wat. We then manually corrected the labeling to ensure bounding boxes precisely enclosed each widget according to rules further detailed in the supplementary materials; notably: when the widget edge was visually identifiable, the bounding box was drawn one pixel away\footnote{Note that differences in annotation padding across datasets mainly affect IoU-based metrics, and barely affect classification metrics such as precision, recall or F1-score.}; when the boundary was ambiguous, the annotator verified the target's extent directly on a real system (\eg by hovering to reveal the interactive region) when possible.
Finally, we assigned class labels according to a fixed taxonomy of six GUI widget types (\textit{ToggleButton}, \textit{TextInput}, \textit{Slider}, \textit{Text}, \textit{HyperLink}, \textit{Button}), further detailed in \autoref{app:gui_element_types}.
Since classification of target \textit{type} is not critical to our envisioned applications, we do not report extensively on this part, but discuss this further in \autoref{sub:limit:classification}.

To evaluate how closely other annotators can reproduce the annotations contained in our dataset while following the same annotation protocol, we randomly sampled 15 images and asked three independent annotators to re-label them using the same guidelines.
We then measured their agreement with the primary annotator as ground truth (see \autoref{app:primary_gt_agreement}), and found high agreement (mean F1 score of 0.933, mean mIoU of 0.892), indicating that both the task and protocol are well-defined, and that the present annotations are reliable.

\begin{figure*}\centering
\includegraphics[width=.85\textwidth]{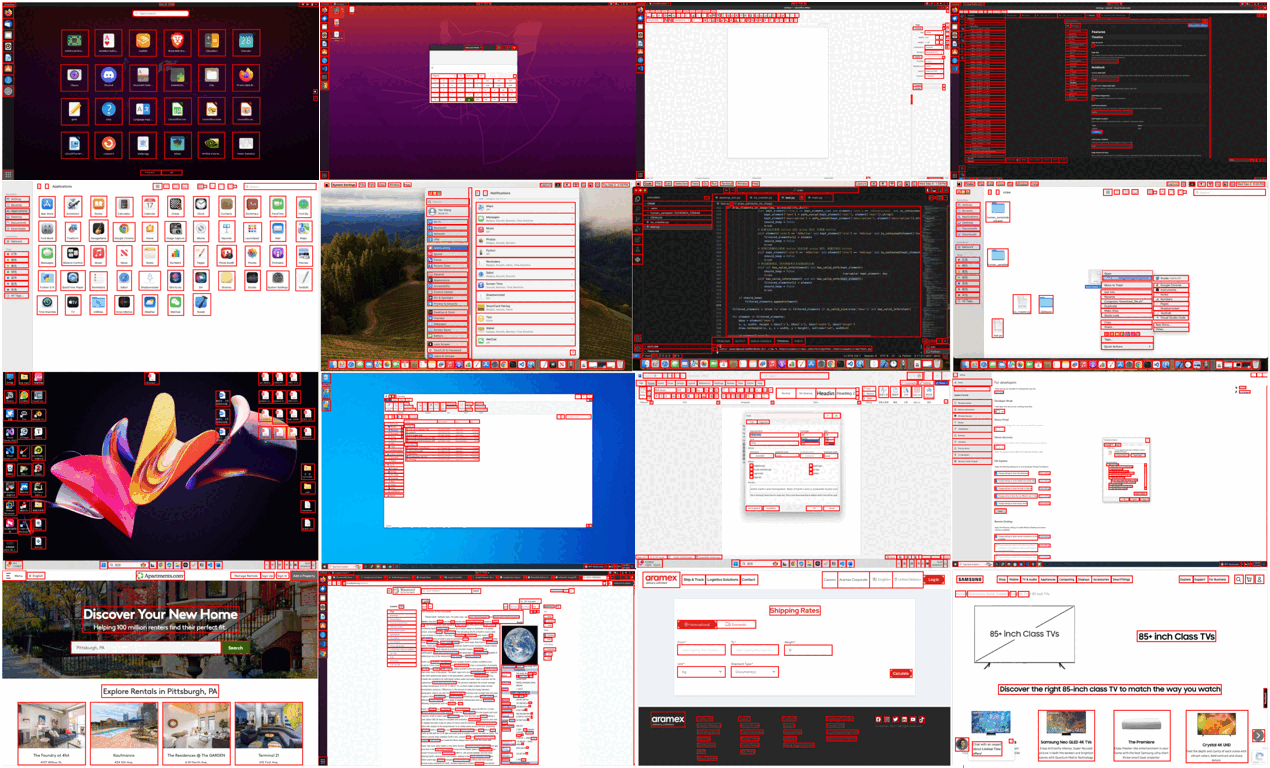}
\caption{Example of hand-labeled screenshots.}
\Description{Example of 16 labeled screenshots from different systems and applications.}
\label{fig:example_labeled_screenshots}
\end{figure*}

\subsection{Dataset presentation}
Our final dataset comprises 520 fully labeled images, evenly distributed across the four platforms: Windows (primarily Windows 10 with some Windows 11), Linux (Ubuntu with GNOME environment), macOS and Web (130 images each).
It includes typical native OS interfaces such as desktop views, system settings, file browsers, calculators, and system tools.
It also covers productivity suites like LibreOffice, Microsoft Word, Excel, PowerPoint, and Draw, as well as developer environments like PyCharm and Visual Studio Code.
For the web category, the screenshots come from different websites such as Amazon, Coursera, Booking, Indeed, and Wikipedia.
A random selection of labeled screenshots is displayed \autoref{fig:example_labeled_screenshots}. The images are predominantly high-resolution; the dominant resolutions for each platform are 2560$\times$1440 for Windows and Linux, 2880$\times$1800 for macOS, and 1440$\times$900 for web interfaces.

\begin{figure}[h]
\centering
\includegraphics[width=.9\columnwidth]{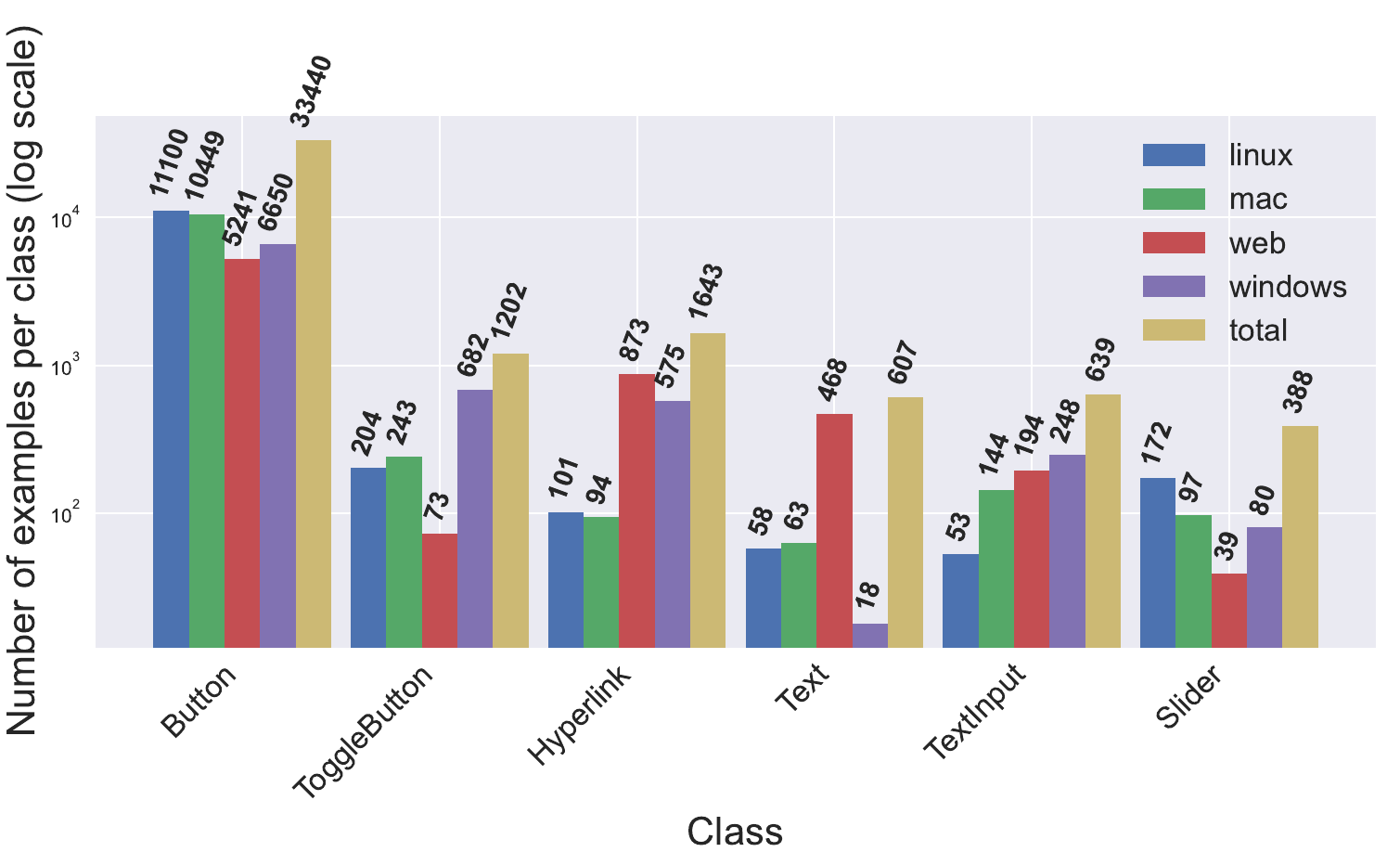}
\caption{Class distribution by category. Note the logarithmic scale on the y-axis.}
\Description{Figure showing the class distribution by category with a logarithmic scale on the y-axis. The categories are Button, ToggleButton, Hyperlink, Text, TextInput, and Slider.}
\label{fig:class-distribution-log}
\end{figure}

Each image is associated with a YOLO-format annotation file.
In total, we annotated 37,919 GUI widgets across six classes.  On average, each image contains approximately 73 widgets, with a minimum of 8 and a maximum of 252 per image.
As shown in \autoref{fig:class-distribution-log}, the dataset exhibits a strong class imbalance, with \textit{Button} elements being the most frequent, particularly on Linux and macOS, while categories such as \textit{TextInput} and \textit{Slider} appear less often; web interfaces contain comparatively more \textit{Text} and \textit{Hyperlink} annotations.
Since we expect that smaller widgets are more difficult to identify, we also report the distribution of widget sizes in our dataset in \autoref{app:yolo_px_analysis}, expressed in YOLO-pixels\footnote{We call YOLO-pixels the geometric dimensions of each widget's bounding box after the screenshot has been rescaled to YOLO’s default input size (640×X). Thus, it is the size that is meaningul in the widget detection context.}.
It reveals an asymetric, positive skewed distribution similar to an EMG distribution~\cite{gori2019}, with a median widget size of about 9 YOLO-pixels.
The dataset\footnote{\url{https://osf.io/fr6y4/overview}} will be made open-access and publicly hosted after the review period.

\section{YOLO training and technical evaluation}

We trained our detection models using the Ultralytics API. The results reported in this section correspond to our main configuration, namely YOLO26n with the default 640$\times$640 input resolution and aspect-ratio-preserving letterboxing. This ensures that screenshots are rescaled without distortion. For comparison with prior work relying on YOLOv8-based detectors, we also report results obtained with a YOLOv8n-640\footnote{The YOLO taxonomy used in this paper is as follows: YOLOXy-z, where $X$ is the model version (\eg v8, 11, 26 \etc), y is the model size (\eg n (nano), s (small), m (medium) \etc), and z is the input resolution (640, 1280, 1920).} model. The effect of varying input resolution and YOLO model size, particularly for the detection of small UI elements and inference speed, is examined separately in \autoref{perfs:tradeoffs}.

\subsection{Dataset preparation and evaluation metrics}
\label{subs:dataset}

\subsubsection{Train/test/validation split} We split the 520 annotated images into three sets: Train\textsubscript{0} (80\%, 416 images) used for hyperparameter tuning, Val\textsubscript{0} (10\%, 52 images) reserved for validation during training, and Test (10\%, 52 images) held out for final evaluation. All splits are stratified to preserve the distribution of GUI classes and platform categories (Windows, macOS, Ubuntu, Web) across subsets.
\subsubsection{Evaluation metrics}
We used standard object detection evaluation metrics (more details are provided in \autoref{app:metrics})
\begin{itemize}
\item \textbf{IoU:} Intersection over Union is used to evaluate a single object detection, and quantifies how much the detected bounding box overlaps with the true bounding box. An IoU of 0.5 means that the overlapping area is 50\% of the union of both boxes.
\item \textbf{F1-score} combines precision (the proportion of predicted boxes that are correct) and recall (the proportion of ground truth boxes that are detected), balancing false positives and false negatives.
\item \textbf{mAP@0.5}: The mean Average Precision at IoU $\geq$ 0.5 measures the model's overall ability to correctly detect objects. It is calculated as the area under the precision-recall curve, counting a prediction as correct when the bounding box has an IoU of at least 0.5 with the ground truth.
\item \textbf{mAP@0.5:0.95}: The mean Average Precision averaged over IoU thresholds from 0.5 to 0.95 is the same except the metric is computed for increasing IoU, in steps of .05 from .5 to .95, and the average over all these mAP is then taken. This evaluates not only the model's ability to find the objects, but also how accurately it places the bounding boxes.
\item \textbf{mIoU} is the mean Intersection over Union computed over all correctly matched detections (true positives).
\end{itemize}
These metrics are all between 0 and 1, with 1 indicating the best performance.
Throughout the paper, we follow the standard YOLO convention that a detection is counted as correct when IoU $\geq$ 0.5.

\subsection{Vanilla training (baseline model)}
To estimate the appropriate number of training epochs, we first conducted a baseline training run on the Train\textsubscript{0}/Val\textsubscript{0} split without any hyperparameter tuning.
We found performance did not show meaningful gains beyond $E=500$ epochs, which served as a reference for subsequent training. The loss and evaluation metrics over the course of training can be found in the supplementary materials.
For {hyperparameter tuning}, we used YOLO's built-in genetic algorithm-based optimizer, running 100 iterations over 150 epochs ($\simeq$E/3).\footnote{We chose 150 epochs as a practical compromise: running each candidate for the full $E=500$ epochs would be time-consuming, since the goal of tuning is to compare many configurations efficiently rather than fully train each one. Shorter runs (\eg 10--50 epochs) would not be sufficient because the model remains too unstable in the early training stages. Using around 150 epochs allows the model to stabilize enough to reveal meaningful differences between hyperparameter sets. For example, with these settings (150 epochs and 100 iterations) on an NVIDIA RTX A6000 GPU, 8-core CPU, and 16 GB RAM of our institution's computing cluster, the process still took approximately 21 hours. Extending the tuning phase over several days (with more iterations) could yield further improvements.}.
This automated search explored both augmentation parameters (\eg mosaic, rotation \etc) and internal YOLO configuration settings, including the classification loss weight, which influences the relative importance of class prediction within the total loss function, which is a key factor in handling class imbalance.

To ensure that the selected hyperparameters did not overfit the initial split,  we performed a 5-fold cross-validation with $E = 500$ epochs. In each fold, 80 \% of the images were used for training and the remaining 20 \% for validation, producing five distinct rotations of the 468 images. The results were consistent across folds, with low standard deviations (see supplementary materials).

\subsection{Final model evaluation and comparison to competing methods\label{sub:comparison_baselines}}
The final model was trained on the full 468-image training set (Train = Train$_0$ + Val$_0$) with the tuned hyperparameters. Its performance was then evaluated on the held-out 52-image Test set for multi-class (\ie bounding boxes with widget classification) and mono-class (bounding boxes only), and compared to the state of the art in widget detection:
REMAUI~\cite{nguyen2015} (the prototype\footnote{\url{http://cseweb.uta.edu/\textasciitilde tuan/REMAUI/}} was originally made available by the authors, but is no longer publicly accessible; we therefore provide our own implementation) , UIED~\cite{xie2020,uied_github}, MobileSAM~\cite{zhang2023} (no public implementation available; we provide our own implementation based on the released model under Ultralytics), \citeauthor{martinez2024}'s instantiation of YOLOv8n-640~\cite{martinez2024,martinez2024_github}, and OmniParser’s YOLO-based detector~\cite{lu2024omniparser}.

OmniParser uses YOLOv8n-640 in the original version presented in their paper, and a larger YOLO11m-1280 model in the more recent release, together with an OCR stage combined with the YOLO detections. Since the OCR stage substantially increases the number of false positives, we report here only the best-performing configuration for object detection, namely the YOLO11m-1280 detector alone (without OCR). A detailed analysis of all OmniParser variants is provided in the supplementary materials.

The results are reported \autoref{tab:perf_final}. Given that the compared methods output a varying number of classes, and some don't return classes at all, comparisons are made based on \textit{mono-class} performance.

\begin{table}[h]\centering
\caption{Performance on the Test set. All performances are single-class evaluations except the one marked (MC). mAPs for REMAUI and UIED are not available since these methods do not return a confidence score associated with each bounding box, which is required to compute mAP scores. The best and best baseline scores are highlighted respectively in blue and light blue.}
\Description{Table showing the performance on the test set for TargetFinder and baseline methods using single-class evaluation, except for YOLO26n-640 (MC), which is multi-class. The reported metrics are Precision, Recall, F1, mAP@0.5, mAP@0.5:0.95, and mIoU. Among the TargetFinder models, YOLO26n-640 achieves the best Precision, Recall, F1, and mAP@0.5, while YOLOv8n-640 achieves the best mAP@0.5:0.95 and mIoU. Among the baselines, OmniParser has the strongest results across all available metrics. mAP values are not available for REMAUI and UIED because these methods do not return confidence scores for bounding boxes.}
\label{tab:perf_final}
\small
\resizebox{\columnwidth}{!}{
\begin{tabular}{@{}lcccccc@{}}
Method & Precision & Recall & F1 & \shortstack{mAP@\\0.5} & \shortstack{mAP@\\0.5:0.95} & mIoU \\
\midrule
\technique & & & & & & \\
\textbf{YOLO26n-640 (MC)}
& 0.769 & 0.695 & 0.723 & 0.738 & 0.510 & 0.866 \\

\textbf{YOLO26n-640}
& \cellcolor{best}{0.936}
& \cellcolor{best}{0.840}
& \cellcolor{best}{0.885}
& \cellcolor{best}{0.899}
& 0.697
& 0.872 \\

YOLOv8n-640
& 0.898
& 0.806
& 0.850
& 0.877
& \cellcolor{best}{0.699}
& \cellcolor{best}{0.880} \\

\midrule
Baselines & & & & & & \\
\shortstack{OmniParser\phantom{aaazzz}\\(YOLO11m-1280)}
& \cellcolor{second}{\raisebox{.75ex}{0.753}}
& \cellcolor{second}{\raisebox{.75ex}{0.651}}
& \cellcolor{second}{\raisebox{.75ex}{0.698}}
& \cellcolor{second}{\raisebox{.75ex}{0.693}}
& \cellcolor{second}{\raisebox{.75ex}{0.391}}
& \cellcolor{second}{\raisebox{.75ex}{0.787}} \\

REMAUI
& 0.430 & 0.484 & 0.455 & N/A & N/A & 0.749 \\

UIED
& 0.397 & 0.476 & 0.433 & N/A & N/A & 0.753 \\

MobileSAM
& 0.348 & 0.167 & 0.226 & 0.063 & 0.030 & 0.767 \\

\shortstack{Martinez-Rojas\\(YOLOv8n-640)}
& \raisebox{.75ex}{0.276} & \raisebox{.75ex}{0.181} & \raisebox{.75ex}{0.218} & \raisebox{.75ex}{0.093} & \raisebox{.75ex}{0.029} & \raisebox{.75ex}{0.684} \\

\end{tabular}
}
\end{table}

\noindent \textbf{Mono-class vs. multi-class:} our model achieves higher scores in mono-class mode (where a true positive is counted if the detection is spatially correct, regardless of the class) compared to multi-class mode (where the predicted class must also be correct), highlighting the impact of class imbalance.\\\textbf{Comparison to other methods:} We found that \technique{} outperformed all baselines, even those using larger YOLO models.
OmniParser is the strongest baseline; it is also the most recent work and its widget detection component was trained on a very large dataset compared to our model. In its best-performing configuration for object detection (YOLO11m-1280 without OCR), it reaches an F1 score of 0.698, compared to 0.885 for TargetFinder with YOLO26n-640 in mono-class mode and 0.850 with YOLOv8n-640. A detailed analysis of all its variants (YOLOv8n-640, YOLO11m-1280, with and without OCR, and OCR+fusion) is provided in the supplementary materials.
Although OmniParser relies on a larger YOLO variant and uses a higher input resolution (1280$\times$1280) than \technique{}, making it about {7.5--9$\times$} slower in our measurements, its performance remains below that of \technique{}, even for very small targets (see \autoref{tab:small_elements_omni} in Appendix~\autoref{other_perfs}).
Omniparser's underwhelming performance can be explained: the system is trained exclusively on $\sim$67k web screenshots annotated automatically from DOM metadata, with no evidence of human verification. Such automatically-generated labels inevitably contain errors (the OmniParser dataset is not public, but the examples shown in~\cite{lu2024omniparser} already reveal labeling errors), and a model trained solely on web UIs does not transfer well to our test set, which includes native desktop environments (Ubuntu, macOS, Windows). This domain mismatch largely explains the lower performance. This observation aligns with the understanding that model performance depends not only on dataset size, but also on the diversity and quality of the data and training, and highlights that large datasets are not always required to obtain good performance.\\

\autoref{fig:mosaic_comparison_labels} illustrates the performances of our trained YOLO model compared with OmniParser's YOLO and the ground truth, on an example screenshot from each OS. Comparison with other methods is provided in \autoref{fig:appendix_mosaic_comparison_labels}.
Our method is visually close to the ground truth, with some errors remaining perceptible (\eg, the bounding boxes not well aligned on the browser suggestions/shortcuts, 1\textsuperscript{st} column) and appears superior to other methods, in line with the technical evaluation.
OmniParser also performs better than the other methods, but still falls short of \technique{}. On native desktop applications it misses numerous widgets (2\textsuperscript{nd} column) and occasionally produces false positives (3\textsuperscript{rd} column).
Its relatively good performance on the browser example (1\textsuperscript{st} column) is expected, since this interface resembles the web domain on which OmniParser was exclusively trained. More detailed evaluations and analyses of all OmniParser variants including on web-screenshots only can be found in the Supplementary materials, which also include other labeled screenshot examples.

{ \setlength{\tabcolsep}{0pt}         \renewcommand{\arraystretch}{0}

\begin{figure*}\centering
\makebox[\textwidth][c]{
\resizebox{0.94\textwidth}{!}{\begin{tabular}{@{} >{\raggedright\arraybackslash}b{0.03\textwidth}
>{\centering\arraybackslash}p{0.35\textwidth}
>{\centering\arraybackslash}p{0.279\textwidth}
>{\centering\arraybackslash}p{0.313\textwidth} @{}}

& \textbf{Ubuntu/Browser example} & \textbf{macOs Example} & \textbf{Windows 10 Example} \\[1pt]

\raisebox{2ex}{\rotatebox{90}{\textbf{Ground truth}}} & \includegraphics[width=\linewidth]{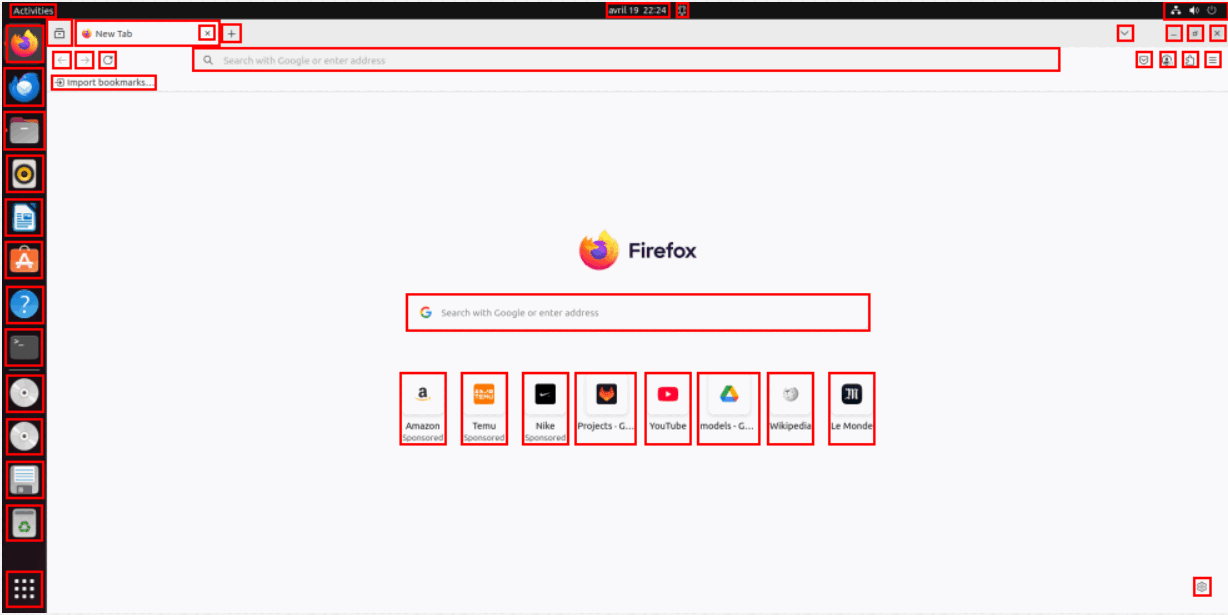}
& \includegraphics[width=\linewidth]{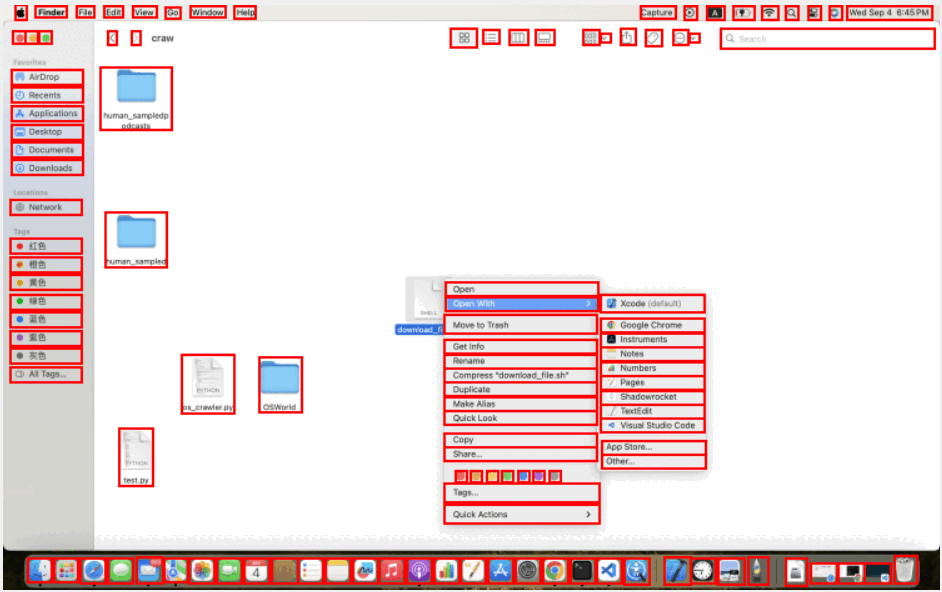}
& \includegraphics[width=\linewidth]{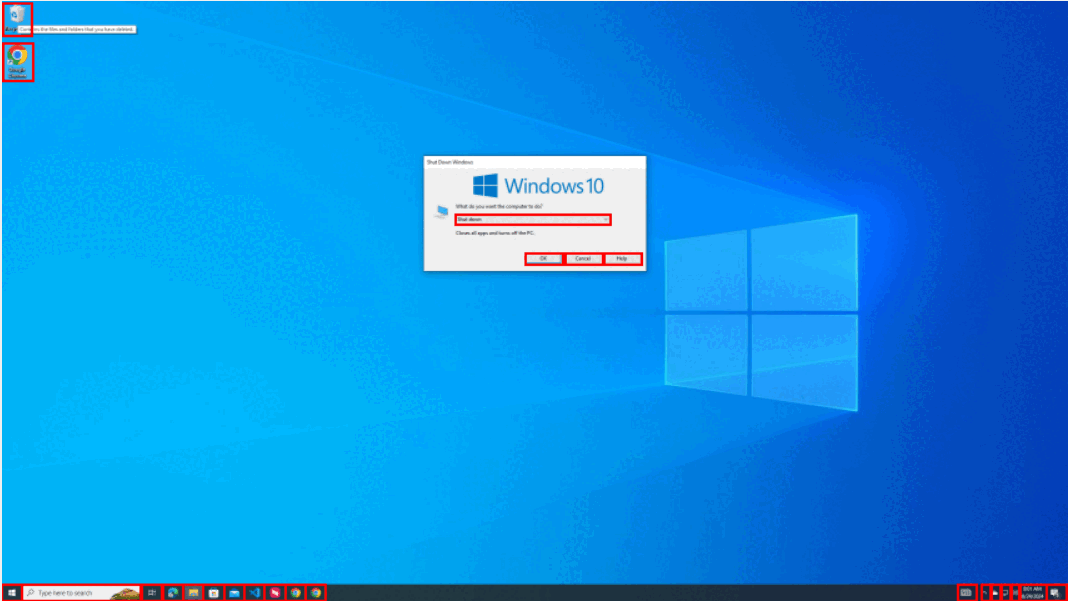} \\[0pt]

\raisebox{2ex}{\rotatebox{90}{\textbf{Our YOLO26}}} & \includegraphics[width=\linewidth]{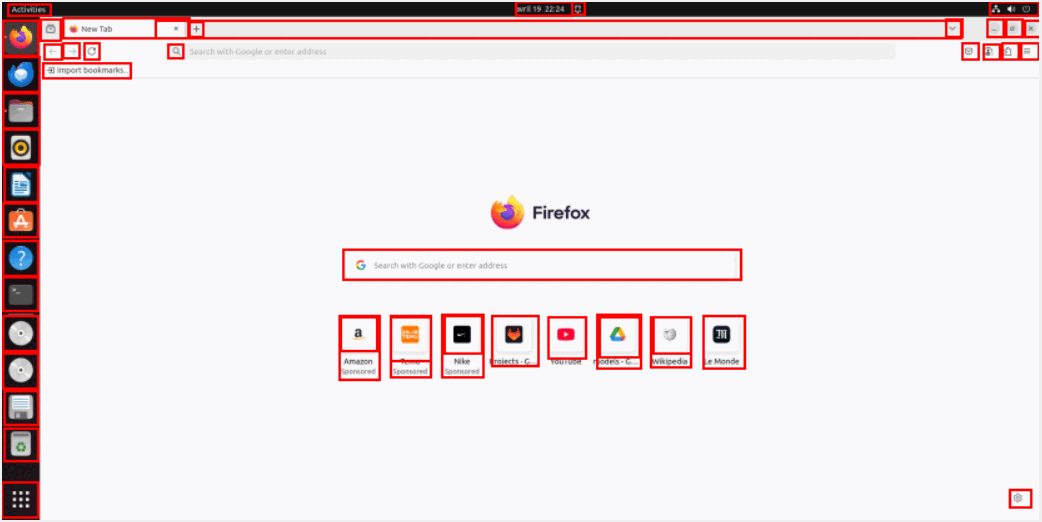}
& \includegraphics[width=\linewidth]{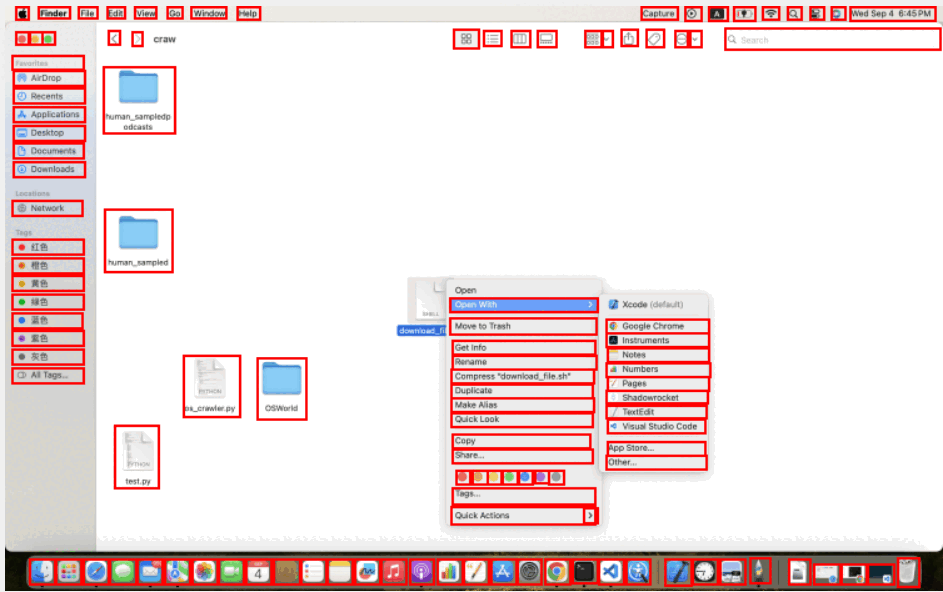}
& \includegraphics[width=\linewidth]{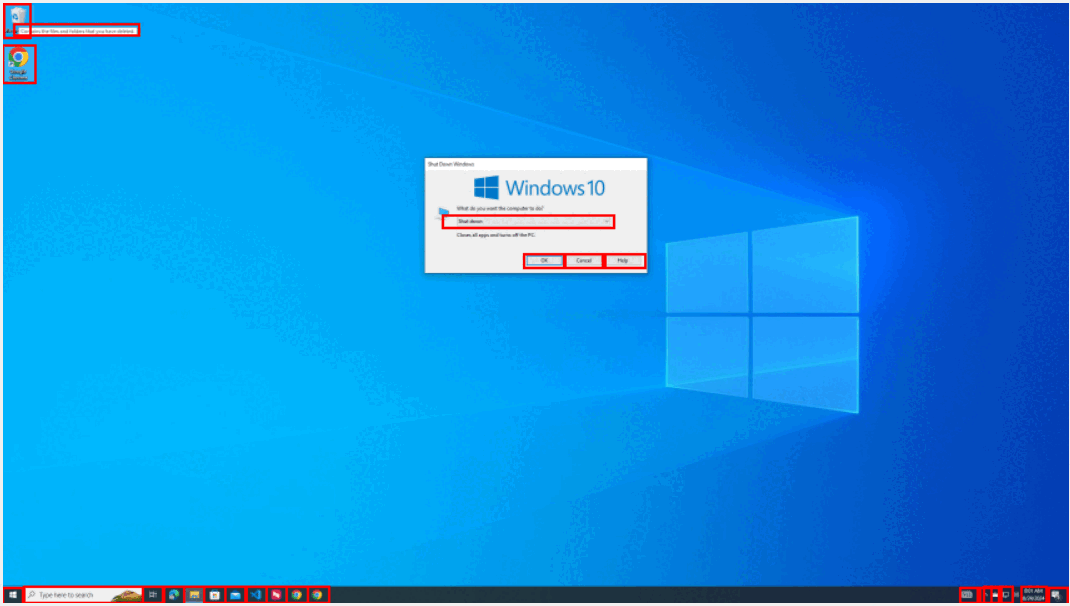} \\[0pt]

\rotatebox{90}{\textbf{OmniParser YOLO}} & \includegraphics[width=\linewidth]{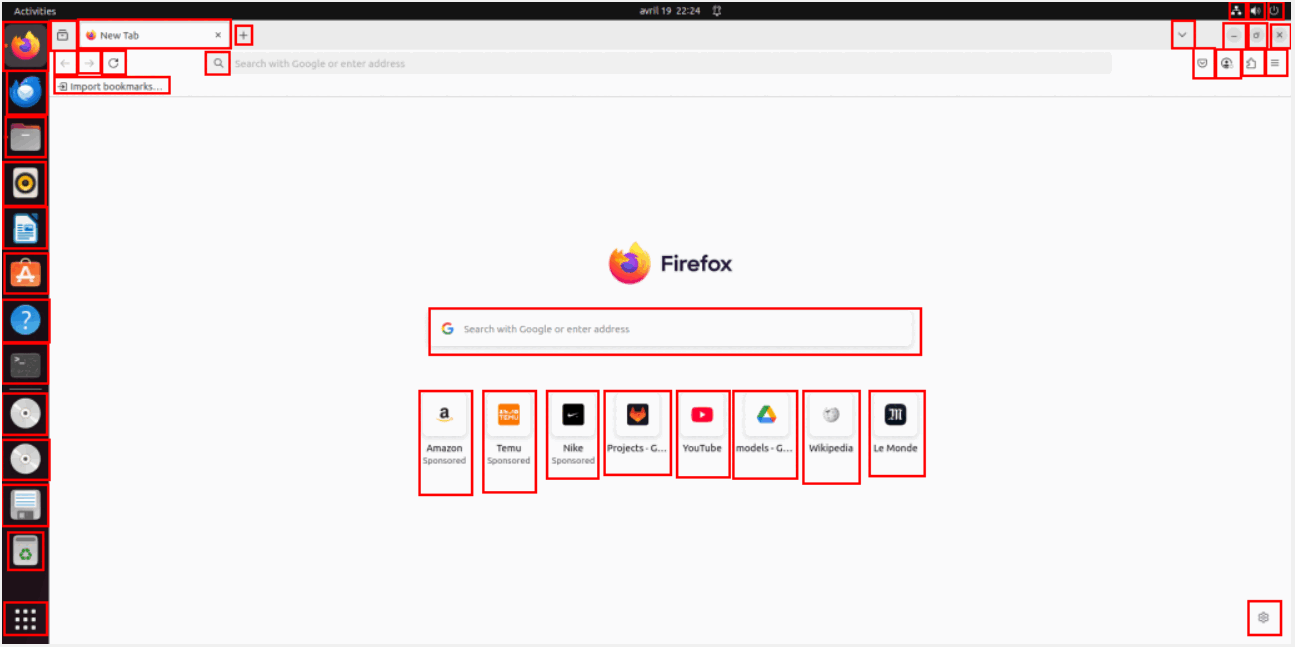}
& \includegraphics[width=\linewidth]{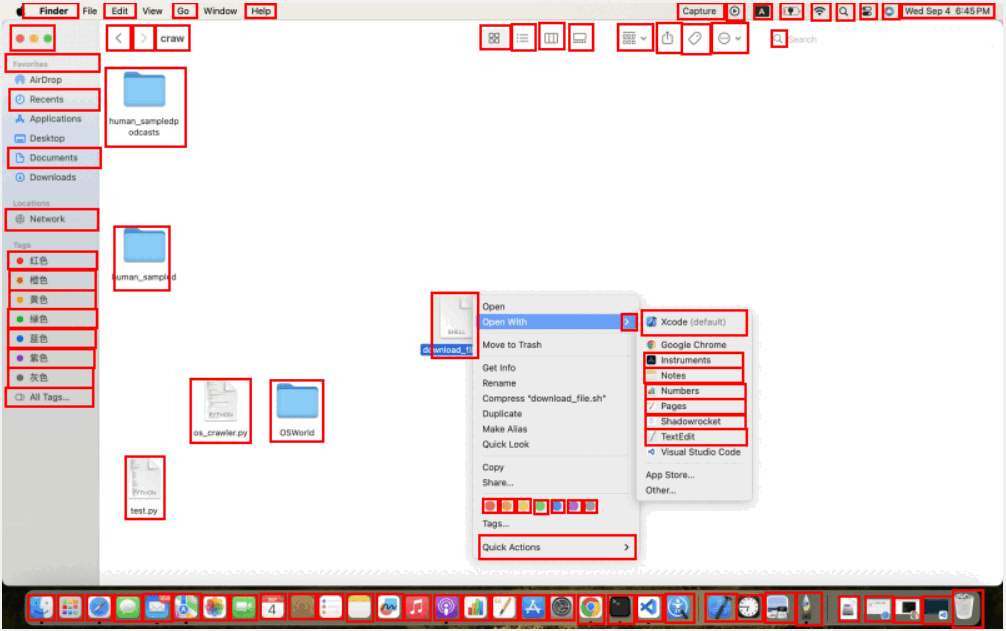}
& \includegraphics[width=\linewidth]{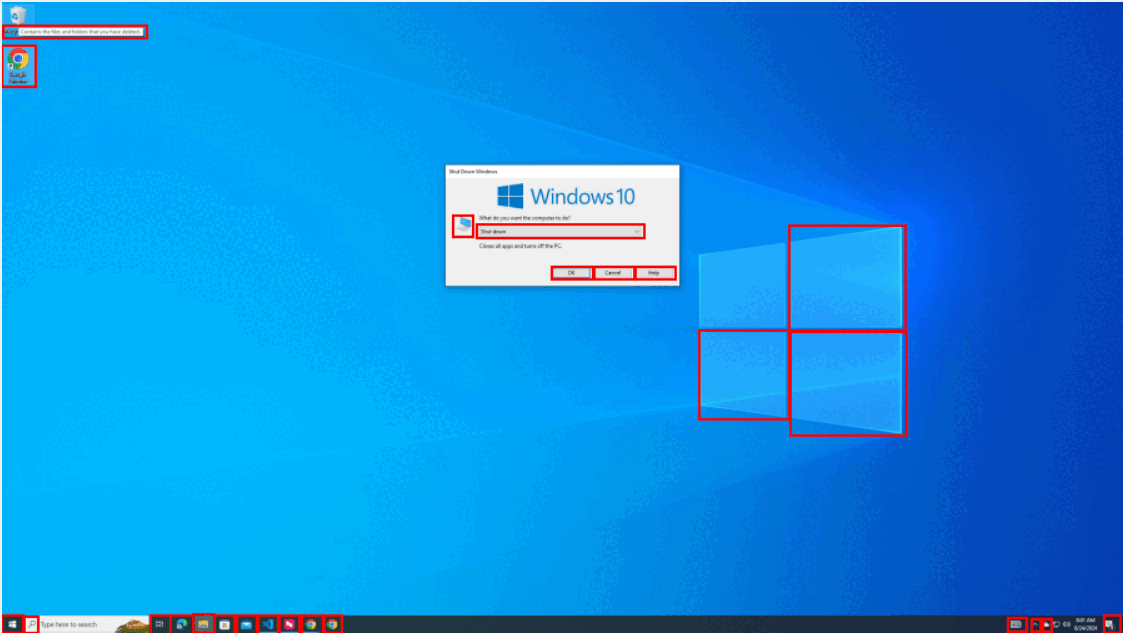} \\[0pt]

\end{tabular}
}
}
\caption{Visual comparison of widget detection outputs from our trained YOLO26n-640 model and OmniParser's YOLO11m-1280 for three example screenshots: a Firefox browser window in Ubuntu (left), a macOs desktop with white background (middle) and a Windows desktop with blue background and a screensaver that can lead to false detections (right).}
\Description{Visual comparison of widget detection outputs from other methods and our trained YOLO26 model for three example screenshots: a Firefox browser window in Ubuntu (left), a macOs desktop with white background (middle) and a Windows desktop with blue background and a screensaver that can lead to false detections (right). The first row shows the ground truth annotations, followed by rows showing the outputs from our YOLO26 model, OmniParser YOLO, REMAUI, UIED, MobileSAM, and Martinez-Rojas YOLO respectively. Each method's output is displayed in a separate row for each of the three examples.}
\label{fig:mosaic_comparison_labels}
\end{figure*}

}

\subsection{Generalization to out-of-distribution data}
Given that our annotated dataset exclusively consists of modern desktop GUI screenshots of three specific OS's, evaluating the final model solely on the corresponding test subset might lead to overly optimistic results.
We thus evaluated the performance of \technique through three experiments: (i) an evaluation on legacy desktop interfaces, which exhibit diverse visual themes and varying image resolutions, (ii) an evaluation on mobile interfaces, and (iii) an evaluation of cross-operating system transfer capabilities, in which models trained on one OS were tested on the two others.

\paragraph{Legacy Software with Different Themes and Resolutions}
We built a test set of 71 manually annotated screenshots from older software covering diverse themes (light, dark) and varying image resolutions, using \wat. The final YOLO26n-640 model, trained on the modern dataset, was evaluated on this legacy set. The results are summarized in~\autoref{tab:generalization_table_performance} under the \textbf{Legacy} column.

\begin{table}[H]
\centering
\caption{Performance on different interfaces (mono-class evaluation).}
\Description{Table showing the performance on different interfaces in a mono-class evaluation. The metrics include Precision, Recall, F1, mAP@0.5, mAP@0.5:0.95, and mIoU for three different test sets: Test-set, Legacy, and Mobile (VINS).}
\label{tab:generalization_table_performance}
\begin{tabular}{lccc}
\toprule
\textbf{Metric} & \textbf{Test-set} & \textbf{Legacy}  & \textbf{Mobile (VINS)}\\
\midrule
Precision & 0.936  & 0.724 & 0.295\\
Recall & 0.840 & 0.669 & 0.347\\
F1 & 0.885 & 0.695 & 0.319\\
mAP@0.5 & 0.899 & 0.717 & 0.201\\
mAP@0.5:0.95 & 0.697 & 0.445 & 0.089\\
mIoU & 0.872 & 0.804 & 0.714\\

\bottomrule
\end{tabular}

\end{table}

The model maintains reasonable performance across precision, recall, and F1, with relative drops of about 21\%. The sharpest decline is observed for mAP@0.5:0.95, with a relative drop of about 36\%, suggesting that performance deteriorates under stricter localization criteria while also reflecting a decrease in overall detection performance, as indicated by the drop in F1. However, the small drop in mIoU (only about 7\%) suggests that, when detections are successfully matched, their spatial overlap with ground-truth objects remains relatively robust.

\paragraph{Mobile Interfaces}
Mobile UIs typically have layouts and design principles that differ significantly from desktop GUIs, so good transfer is a priori not expected. Still, given that desktop interfaces are generally more complex than mobile ones, it is plausible that transfer is possible, but only in the desktop-to-mobile direction.
We thus evaluated the model on 100 mobile screenshots (Android and iPhone) sampled from the VINS dataset, reported under \textbf{Mobile (VINS)} in \autoref{tab:generalization_table_performance}.
The drop in performance confirms that desktop-trained models do not generalize well to mobile UIs. Training on mobile data is thus essential to ensure reliable performance in mobile contexts.

\paragraph{Cross-OS Transfer}
A practical question of interest is whether the detection results transfer across modern interface themes; indeed, modern interfaces tend to share many similarities since they are often based on the same heuristics of design\footnote{See \eg Jakob Nielsen's 10 heuristics of design: \url{https://www.nngroup.com/articles/ten-usability-heuristics/}.}. To investigate this, we trained YOLO26n-640 models individually on the Windows-, Ubuntu-, and macOS-only images of our dataset, and evaluated their performance across the other OSs.
Each OS-specific dataset comprises 130 images, divided into 80\% training, 10\% validation, and 10\% test subsets. All training used the default YOLO hyperparameters pre-configured for the COCO dataset~\cite{lin2014}, with a maximum of 500 epochs,
providing a consistent baseline for comparison.

\begin{table}\centering
\caption{Cross-OS transfer performance (mono-class evaluation).}
\Description{Table showing the cross-OS transfer performance in a mono-class evaluation. The metrics include Precision, Recall, F1-score, mAP@0.5, mAP@0.5:0.95, and mIoU for different training and testing OS combinations: Windows-Windows, Windows-Ubuntu, Windows-macOS, Ubuntu-Ubuntu, Ubuntu-Windows, Ubuntu-macOS, macOS-macOS, macOS-Ubuntu, and macOS-Windows.}
\label{tab:os_transfer}
\small
\resizebox{\columnwidth}{!}{
\begin{tabular}{@{}llcccccc@{}}
\toprule
\textbf{Training} & \textbf{Testing} & \textbf{Precision} & \textbf{Recall} & \textbf{F1-score} & \parbox{1cm}{\centering\textbf{mAP@\\0.5}} & \parbox{1cm}{\centering\textbf{mAP@\\0.5:0.95}} & \textbf{mIoU} \\
\midrule
Windows & Windows & \cellcolor{best}{0.807} & 0.643 & 0.716 & 0.725 & \cellcolor{best}{0.473} & \cellcolor{best}{0.827} \\
Windows & Ubuntu & 0.666 & 0.576 & 0.618 & 0.634 & 0.324 & 0.751 \\
Windows & macOS & 0.801 & \cellcolor{best}{0.710} & \cellcolor{best}{0.753} & \cellcolor{best}{0.779} & 0.350 & 0.736 \\
\midrule
Ubuntu & Ubuntu & \cellcolor{best}{0.868} & \cellcolor{best}{0.791} & \cellcolor{best}{0.828} & \cellcolor{best}{0.852} & \cellcolor{best}{0.561} & \cellcolor{best}{0.819} \\
Ubuntu & Windows & 0.472 & 0.362 & 0.410 & 0.339 & 0.130 & 0.722 \\
Ubuntu & macOS & 0.645 & 0.670 & 0.657 & 0.644 & 0.303 & 0.751 \\
\midrule
macOS & macOS & \cellcolor{best}{0.955} & \cellcolor{best}{0.931} & \cellcolor{best}{0.943} & \cellcolor{best}{0.973} & \cellcolor{best}{0.802} & \cellcolor{best}{0.901} \\
macOS & Ubuntu & 0.561 & 0.555 & 0.558 & 0.525 & 0.199 & 0.696 \\
macOS & Windows & 0.571 & 0.380 & 0.457 & 0.387 & 0.131 & 0.689 \\
\bottomrule
\end{tabular}
}

\end{table}

The results, summarized in \autoref{tab:os_transfer}, show that cross-OS transfer clearly exhibits performance declines, particularly when transferring models from macOS or Linux to Windows.
Overall, performance is highest for the MacOS(train)/MacOS(test) pair, suggesting that the MacOS interface exhibits less diversity across our dataset.
Finally, in many cases, F1 scores and mIoU remain decent; this suggests that, while models trained on a single OS can partially generalize to other platforms for some practical applications, fine-tuning or additional retraining with target-specific data will enhance results significantly.

\section{Identifying Small widgets\label{sec:small_widgets}}
Small widgets are common in GUI's, including in our dataset (see \autoref{fig:yolo_px_distribution}). They are usually hard to detect, and small object detection remains an active research topic~\cite{wang2021}. In this section, we first characterize the prevalence of small widgets in our dataset, and then apply and evaluate several methods to enhance their detection.

\subsection{Distribution of widget sizes in our dataset and performance on small elements}
\label{app:yolo_px_analysis}

\begin{figure}[h]\centering
\includegraphics[width=0.9\linewidth]{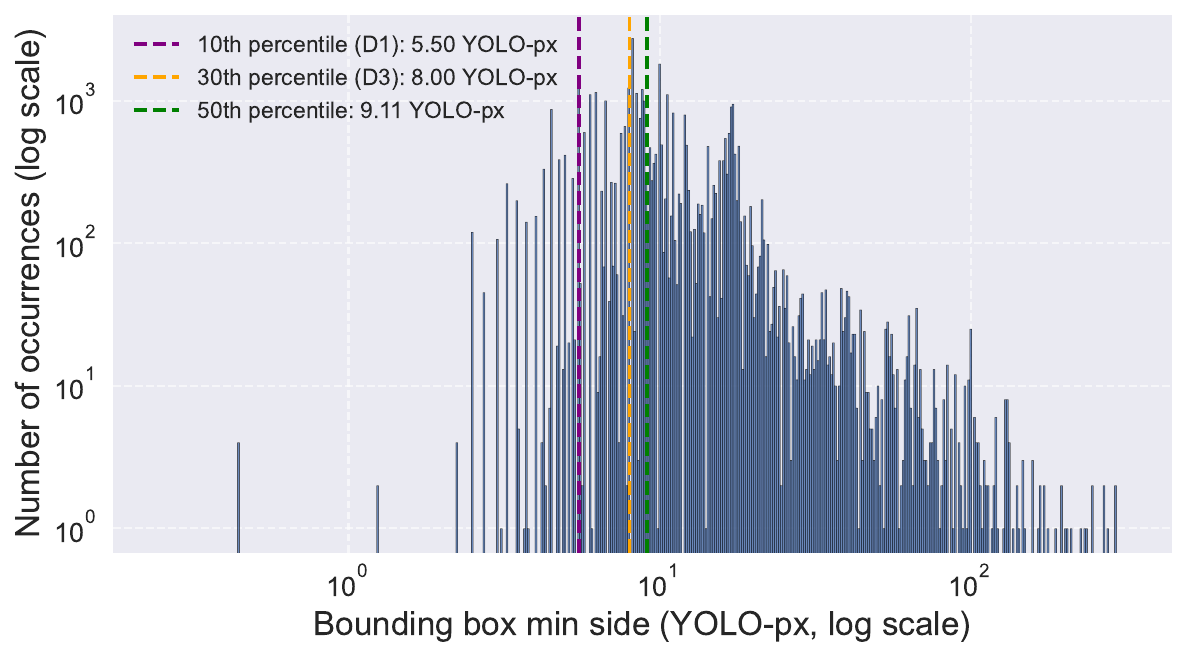}
\caption{Distribution of widget sizes in YOLO-pixels (logarithmic scale), when a network with input resolution of 640$\times$640 is used. Vertical lines indicate the 50th, 30th, and 10th percentiles.}
\Description{Figure showing the distribution of widget sizes in the dataset measured in YOLO-pixels after letterboxing. The x-axis represents the minimum side length of each bounding box on a logarithmic scale. Vertical lines indicate the 50th, 30th, and 10th percentiles of the size distribution.}
\label{fig:yolo_px_distribution}
\end{figure}

To characterize the presence of small UI elements in our dataset, we computed the distribution of widget sizes \emph{after} letterboxing as the minimum between widget width and height, expressed in ``YOLO-pixels''\footnote{After resizing each screenshot to 640$\times$X while preserving aspect ratio, letterboxing adds padding to obtain the final 640$\times$640 input resolution. “YOLO-pixels” do not correspond to physical image pixels, but represent the bounding box's dimension \emph{after} resizing. This implies that YOLO-pixels are floats.}, see \autoref{fig:yolo_px_distribution}.

The performance of \technique was evaluated separately for widgets whose YOLO-pixel size falls below the 50th, 30th, and 10th percentiles (see ~\autoref{tab:small_elements_640} in Appendix~\ref{other_perfs}), and found that, in line with the literature, smaller widgets were increasingly difficult to detect, as reflected by the higher false-negative rates (lower recall).  The best baseline, OmniParser, suffers from a similar decrease of recall (see \autoref{tab:small_elements_omni} in Appendix~\autoref{other_perfs}). REMAUI also exhibits a clear drop in performance across percentiles (see \autoref{tab:small_elements_remaui} in Appendix~\ref{other_perfs}), although in this case the degradation appears to affect precision more strongly than recall.

\subsection{Enhancing performance for small widgets}
We evaluated several methods for improving performance for small widgets. The performances of all relevant YOLO variations, evaluated on the $10^{\text{th}}$ percentile and on mAP@0.5:0.95, are provided in \autoref{fig:yolo_speed_vs_map_10th_perc}.
\begin{enumerate}
\item \textbf{Specific small-widget training} brings marginal to no improvements. We experimented with several fine tunings of hyperparameters and YOLO augmentations, as well as training with a curated dataset of images containing many small widgets, but found almost no gain in performance, as reported in the Supplementary materials. The problem is that small targets become almost invisible, deep in the CNN-backbone of the YOLO network. For example, a widget of size 5.5 YOLO-pixels becomes sub-pixel in deeper feature maps ($\approx $0.34 pixels at stride 16).
\item \textbf{Specific YOLO architecture.} The YOLO family offers various model sizes, denoted n (nano), s (small), m (medium), l (large), and x (extra-large). We fine-tuned models of different sizes and found sizeable gains in detection performance, but at the cost of inference time.
Other changes to the YOLO architecture have been reported in the literature.
For example, YOLO uses information from multiple convolutional layers for detection (feature pyramids~\cite{lin2017}). YOLO does not use the P2 layer, which is the layer with the highest resolution \ie the one most suited to capture information related to small widgets. We experimented with an undocumented YOLO architecture that exploits the P2 layer, but found at best marginal improvements, reported in the supplementary materials.\footnote{The untrained model can be found at \url{https://github.com/ultralytics/ultralytics/blob/main/ultralytics/cfg/models/26/yolo26-p2.yaml}. Previous work shows that adding P2 can bring small improvements to detection, \eg an increase of .04 points in MAP@0.5:0.95 in~\cite[Table 3]{qiu2025}, but requires training YOLO from scratch. An alternative, which we considered instead, is to transfer the weights from a previously pre-trained YOLO without P2 and learn the remaining weights by fine-tuning.} Other works also suggest changing the loss function during training, moving away from IoU since it becomes very sensitive for small targets, but reported improvements appear even smaller~\cite[Table 4]{qiu2025}.
\item \textbf{Increasing the input resolution} is a simple yet very effective method to increase performance on small widgets: doubling the resolution doubles the size of the objects represented inside the CNN-backbone. We provide YOLO models trained and evaluated at different input resolutions, including 640$\times$640, 1280$\times$1280, 1920$\times$1920.
\item In \textbf{Slicing Aided Hyper Inference (SAHI)} the image is cut into overlapping patches. Each patch is fed to a YOLO network, after which the detection results are fused. The advantage is that downscaling the initial image is not necessary anymore, and the extra time needed to run the extra forward inference passes on YOLO can, in theory, be negated by running these in parallel. We indeed found a gain in performance, but at a higher temporal cost than expected. We gained further performance by training the networks on patches rather than on full images.
\end{enumerate}

\subsection{Performance tradeoffs}\label{perfs:tradeoffs}
\begin{figure}[htbp]
\centering
\includegraphics[width=\columnwidth]{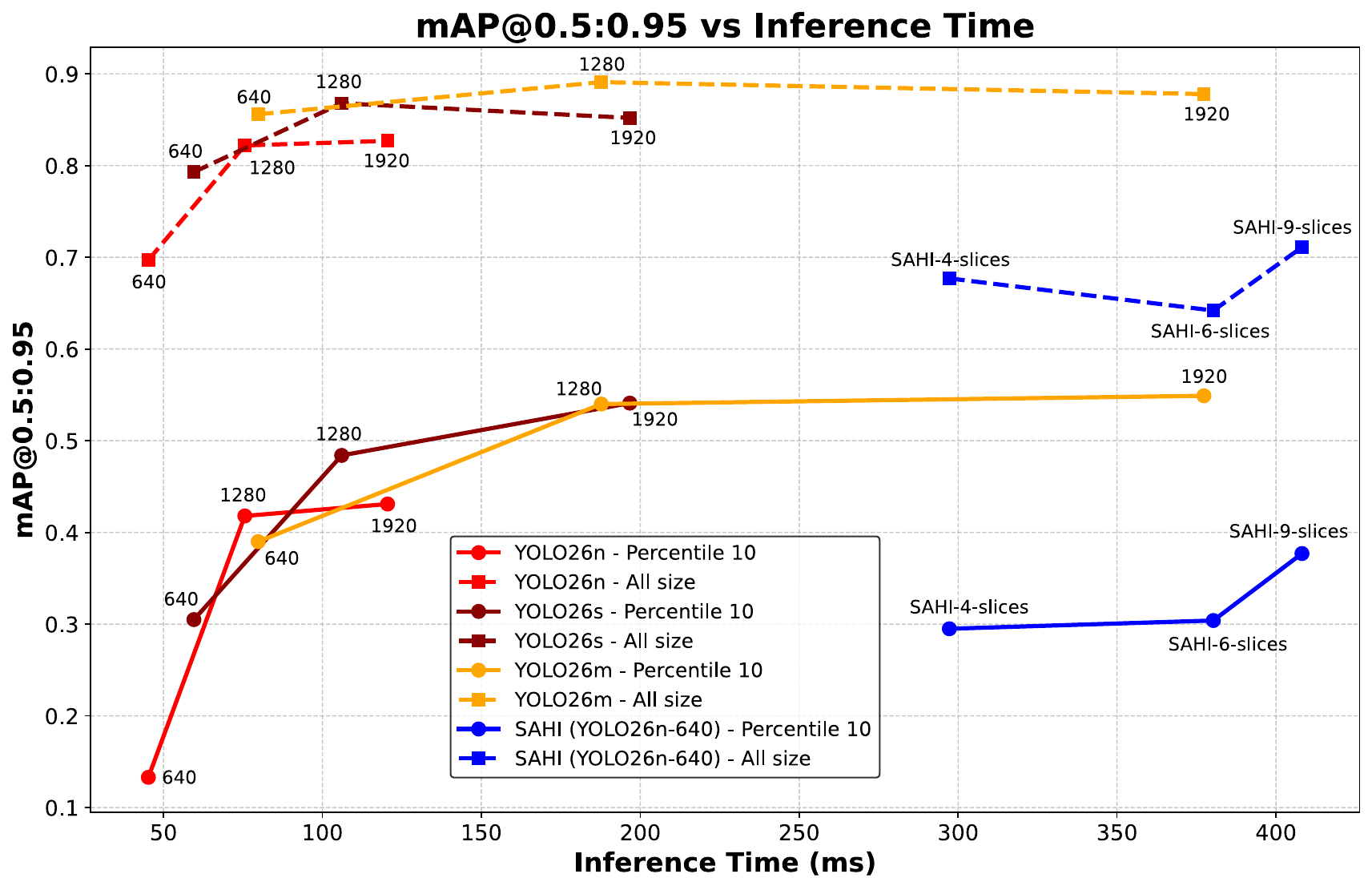}
\caption{Performance comparison for different YOLO architectures for small targets (10th percentile). The reported SAHI benchmark uses YOLO26n-640. Inference time is given for a Macbook Pro (2023) with M3 Max chipset (14 cores), 36 GB RAM, no dedicated GPU.}
\Description{Line chart comparing mAP@0.5:0.95 and inference time for several YOLO26 models and SAHI with YOLO26n-640. The x-axis shows inference time in milliseconds, and the y-axis shows mAP@0.5:0.95. Results are reported for both all widget sizes and small widgets at the 10th percentile. Larger models and higher input resolutions generally improve accuracy at the cost of slower inference, while SAHI provides some accuracy gains but with a large increase in inference time.}
\label{fig:yolo_speed_vs_map_10th_perc}
\end{figure}

Our evaluations of the different YOLO models reveal that 1) larger models (parameters and input resolution) deliver higher accuracy at the expense of slower inference speeds 2) SAHI provides some gains in performance but for a large increase in inference time 3) A variety of other methods can bring marginal gains in performance.
From a practical perspective, the choice of a particular version of the model should be made based on the inference speed requirements. For the use cases presented next, a 200\,ms inference time is acceptable, which, with the hardware used to create the \autoref{fig:yolo_speed_vs_map_10th_perc} benchmark, suggests using the $s\times1920$ model. The supplementary materials reports benchmarks for different hardware setups, including high- to low-end CPUs, and a consumer-grade GPU.

\section{\technique}
\subsection{System description and technical evaluation}
\technique continuously captures full-screen screenshots at 30Hz\footnote{This parameter can be increased on powerful machines.} with a cross-platform screenshot library optimized for speed called \texttt{mss}\footnote{\url{https://pypi.org/project/mss/}}.
To avoid redundant YOLO inferences when the UI remains static and save computational resources, \technique integrates a change detection mechanism. After each screen capture, a downscaled grayscale version of the current frame is compared to the previous one. If the pixel difference exceeds a configurable threshold, a new inference is triggered using our trained model; otherwise, the inference step is skipped.

YOLO inference is configurable through confidence and IoU thresholds. The default confidence threshold is chosen to maximize the F1-score, but can be adjusted to favor precision (fewer false positives but more missed detections) or recall (fewer missed detections but more false positives).
The IoU threshold is also adjustable: lower values eliminate overlapping boxes more to reduce duplicates via non-maximum suppression.

The detection runs in a separate thread, ensuring that bounding boxes are updated in real time. Each detection is primarily represented by a dictionary \texttt{\{id, x, y, width, height, score, class\_id, class\_name\}}\footnote{\texttt{x} and \texttt{y} are the top-left coordinates of the bounding box; \texttt{w} and \texttt{h} its width and height; \texttt{score} the confidence in $[0,1]$; and \texttt{class\_id} the class label.}, ready to be exploited by downstream modules. Optionally, the library can also provide the cropped image of each detected widget, which can be useful for further processing such as OCR or custom classification.
In practice, the speed of \technique depends mainly on the YOLO inference latency and on the cost of the change detection operation.
On a Macbook Pro 2023, with a M3 Max chipset (14 cores), 36~GB RAM, the system reaches 10--24~FPS for real-time detection, depending on whether the interface is static or dynamically changing, with CPU usage ranging from ~4\% (static UI) to ~7\% (dynamic inference) and RAM usage 390--470~MB.

\captionsetup[subfigure]{justification=centering,singlelinecheck=false}
\begin{figure*}[htp]
\centering
\begin{subfigure}{0.375\textwidth}
\includegraphics[width=\linewidth]{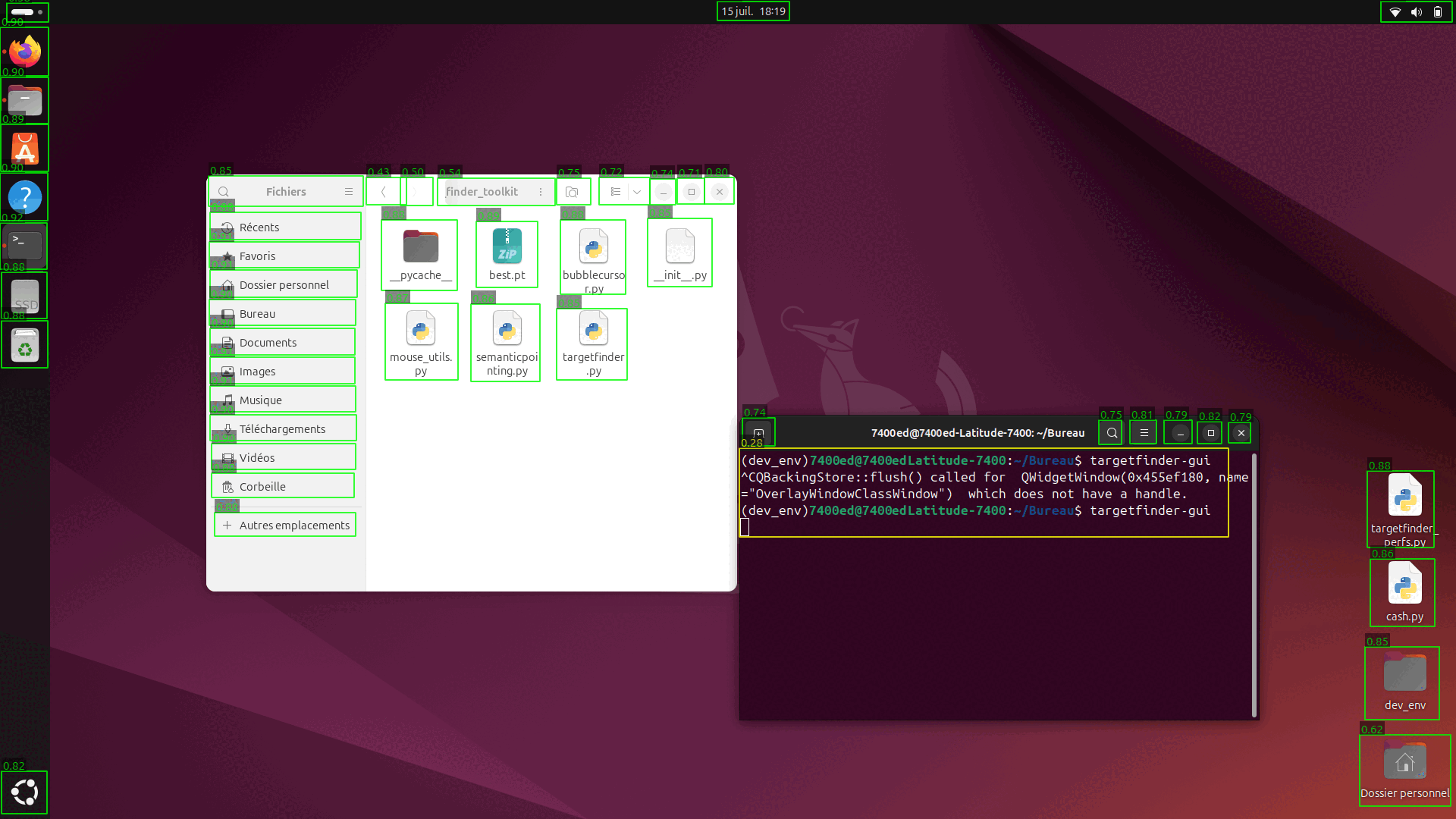}
\caption{Example of detected widgets\\ by \technique.}
\end{subfigure}
\hfill
\begin{subfigure}{0.305\textwidth}
\centering
\includegraphics[width=\linewidth]{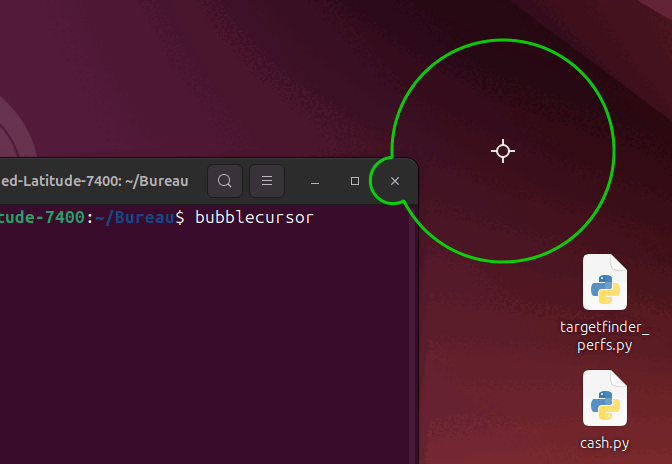}
\caption{Example of the Bubble Cursor overlay\\ during real-time use.}
\end{subfigure}
\hspace{-0.8em}
\begin{subfigure}{0.3\textwidth}
\centering
\includegraphics[width=.8\columnwidth, trim = {1cm 2.1cm 1cm 1.5cm}, clip]{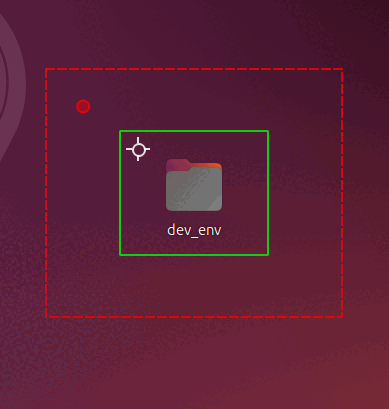}
\caption{Example of semantic pointing with visual\\ feedback about the motor space displayed.}\end{subfigure}

\caption{\technique and system-wide implementations of BubbleCursor and SemanticPointing illustrated.}
\Description{(a) Example of detected widgets by TargetFinder. The detected widgets are highlighted with green bounding boxes. (b) Example of the Bubble Cursor overlay during real-time use. The current nearest target is the close button of the terminal window. It is represented by a circle centered on the cursor and extending toward the center of the identified widget, merged with a rounded rectangle around the widget. (c) Example of semantic pointing with visual feedback about the motor space, where the target is outlined in green in visual space and the enlarged motor space is shown with a red dashed boundary.}
\label{fig:bubble}
\end{figure*}

\subsection{Toolkit\label{sec:toolkit}}
The \technique library, provided as supplementary material, manages screen capture, low-resolution change detection, and YOLO inference. The library can be imported into any Python application to retrieve real-time bounding boxes of visible interface widgets. It exposes configurable parameters such as loop frequency, change-detection threshold, YOLO confidence, and IoU threshold for non-max suppression. Upon detection of a change, the library provides a callback mechanism to notify the application, providing information about the detected changes (id, position, dimensions, class, confidence score and screen capture of each widget).
We also provide all fine-tuned YOLO models considered in this work, allowing end-users to navigate the mAP@0.5:0.95/inference time tradeoff.
The detection pipeline is designed to be platform-agnostic. However, overlay rendering using Qt6 and mouse control (used in Bubble Cursor and Semantic Pointing) rely on system-level APIs that differ between operating systems. These components have been validated on Windows and Ubuntu (X11 sessions). Wayland\footnote{\url{https://wiki.archlinux.org/title/Wayland}} is not currently supported due to limitations in available screen capture libraries, and adaptations are necessary on macOS.

We now present two demonstrations of the functionality enabled by \technique, showing how it allows a system-wide implementation of two existing target-aware pointing facilitation techniques. The techniques are illustrated on an Ubuntu 24.04 system. Videos of the systems working in real-time and system-wide are available in the supplementary materials, as well as further details on the technical solutions used.

\subsection{Illustration in bubble cursor}
We implemented the Bubble Cursor using \technique, following the algorithm described by \citeauthor{grossman2005}~\cite{grossman2005} which continuously adjusts the cursor's radius to select only the closest target inside the active area.
To display the bubble cursor (see \autoref{fig:bubble}-b for an example), we use a PyQt overlay while completely hiding the system cursor using low-level system functions\footnote{We used the Python ctypes library. Although PyQt can hide the system cursor, in practice, due to conflicts between screen capture, overlay refresh, and click handling, we chose to hide the system cursor entirely for robustness. More details on our engineering choices are provided in the supplementary materials.}. This overlay is updated every 10\,ms.
When the user clicks, the system automatically redirects the hidden pointer to the center of the nearest widget using the \texttt{pyautogui}\footnote{\url{https://pypi.org/project/PyAutoGUI/}} library to simulate a click, then moves the pointer back to its original location to ensure that the interaction remains
smooth.

\subsection{Illustration in semantic pointing}
Semantic pointing~\cite{blanch2004} dynamically adjusts the control-display ratio~\cite{casiez2011} (C/D), based on how close the cursor is to nearby targets. Decreasing the C/D ratio when the cursor is near or within a target causes the cursor to slow down, meaning the user must move their hand farther to traverse the same number of pixels, which enlarges the target in motor space while keeping its appearance in visual space unchanged.
To implement Semantic pointing, as in the Bubble Cursor implementation, we fully hide the system cursor and render a fake cursor using PyQt, refreshing every 10\,ms.  We apply the spatial scaling function as defined in the original work by \citeauthor{blanch2004}~\cite{blanch2004}, parametrized by the semantic importance \(S_i\):
for \(S_i < 1\), the cursor moves faster, and the widget appears smaller in motor space; conversely for $S > 1$, the cursor moves slower and the widget appears larger in motor space. \autoref{fig:bubble}-c illustrates\footnote{This additional visual layer can be displayed with the \texttt{--display} option in our implementation, but by default the size in motor space is hidden, and the technique is visually imperceptible to users.} the bounding box of a target in visual space as determined by \technique (green), and the associated size in motor space with the semantic pointing technique (red, dashed) for $S_i=2$.

\section{Limitations}

\paragraph{Limited number of widget classes\label{sub:limit:classification}} We considered 6 classes in the current implementation of \technique, but prior work has generally used more, \eg 27 in~\cite{martinez2024}.
Considering that our intended applications do not actually require knowing the type of widget being targeted, and given that performance comparisons with existing techniques are necessarily mono-class, we opted for a limited number of high-level classes. It should also be noted that a universal set of classes does not exist: for example, OS-Atlas uses 34 classes for Linux and 24 for Mac OS~\cite{wu2024}.
Another example is in \autoref{fig:hyperlink_text}, where it is not obvious whether the \textit{DatePicker} widget should be classified as \textit{Button}, a \textit{Text field}, or a class on its own.
Further, the choice of classes is dependent on the application; for example the authors of semantic pointing describe how they reduced the motor space of the [\textit{Don't save}] button but increased the motor space of the [\textit{Save}] button. Thus, a class-aware version of semantic pointing would not consider a \textit{Button} class
so much as classes of recovery-from-error costs.
We addressed this limitation in two ways: first, we elected to facilitate the relabeling of widget classes as much
as possible by releasing our annotation tool, which is easily extendable to
more classes. Second, \technique optionally provides a screenshot of each widget alongside geometric information, allowing further downstream classification \eg based on OCR~\cite{li2023}.

\paragraph{Limitations of \technique in practice}

\technique currently does not identify hierarchies of widgets, such as the fact that the close tab icon belongs to the tab button, nor does it return entire windows, which are sometimes the target of users wanting to deselect or switch focus. These cases are likely better treated with extra post-processing, rather than by adding specific classes such as \textit{window} or \textit{compounded widget}.
For instance, if a click occurred on an identified target, re-running YOLO within the bounding box of the widget to see if other sub-widgets are found within; and if not, analyzing the whole image looking specifically for windows.

\paragraph{Dataset size}
Our dataset contains more than 42,000 annotations from about 600 screenshots, far from OmniParser's training set, which has more than 60k screenshots. Yet, thanks to manual labeling and inclusion of desktop screenshots, \technique outperforms OmniParser. This result reinforces that the \textit{quality} of the data is essential~\cite{halevy2009, sutton2019}. That being said, further performance enhancements can likely be gained by increasing the size of training set.

\paragraph{Real-time performance\label{sub:limit:realtime}}
The end-to-end latency of our system is about 200\,ms.
While this may seem like a large delay, considering the effect that latency has on pointing performance when evaluated by Fitts' law~\cite{hoffmann1992}, this processing delay is not equivalent to adding a constant latency to the pointer: it only affects the brief moment immediately after an interface change (e.g., when a new window is opened), during which the widget information cannot be exploited.
Still, since inference time can be traded-off for precision and recall (see \autoref{sec:small_widgets}), it remains an important concern.

\section{Conclusion and Future Work}
This work introduces \technique, a real-time, computer vision-based library for identifying GUI widgets across desktop environments. Unlike traditional approaches that rely on accessibility APIs or application-specific access, \technique operates independently of platform internals, enabling system-wide widget detection.
To support this, we compiled a novel dataset of 520 annotated screenshots spanning multiple OSes and interface types. We fine-tuned multiple YOLO object detection models, of various sizes and input resolution, on this dataset, achieving significantly higher widget detection accuracy compared to existing methods. Finally, we illustrate the immediate applicability of \technique by enabling two classic target-aware pointing techniques, Bubble Cursor and Semantic Pointing, as drop-in enhancements on standard Windows systems.
Each of these contributions (annotation tool, annotated dataset, trained models, \technique and pointing techniques code) are made publicly available.
These contributions pave the way for wider deployment of advanced interaction techniques without requiring cooperation from application developers or access to source code.
Our findings also shed light on the generalization capabilities of object detection methods in desktop GUIs, showing the importance of including images from various OS's and themes in the training set.
Apart from building target-aware interaction techniques, we see two other avenues for future work with \technique:

\begin{itemize}
\item{Ecological} (or ``in the wild'') {studies of interaction}, that evaluate pointing behavior in
realistic environments. \technique should facilitate the validation of established models like Fitts' Law in real-world contexts~\cite{chapuis2007}, ecological comparisons between interaction techniques, and tuning of pointing facilitation mechanisms~\cite{lee2020}. \item As highlighted \autoref{sub:rw:alternate_methods}, many applications implement {accessibility} APIs incompletely or incorrectly, creating barriers for assistive technologies. Reliable spatial and semantic information about interface widgets provided by tools such as \technique could contribute to more robust screen readers, magnification tools, or alternative input methods that can operate independently of developer-provided accessibility metadata.
\end{itemize}

\begin{acks}
\finalonly{Thanks to Fabio Matti, Ludovico Novelli and Rosie Frost for insightful private discussions. This work was partially funded by ANR-24-CE33-7994-01 HCIMI.}
\end{acks}

\bibliographystyle{ACM-Reference-Format}
\bibliography{targetfinder}

\section{Appendices}

\subsection{Semi-automatic annotation tool}
\label{fig:annotation_tool}
\begin{figure}[H]
\centering
\includegraphics[width=\columnwidth]{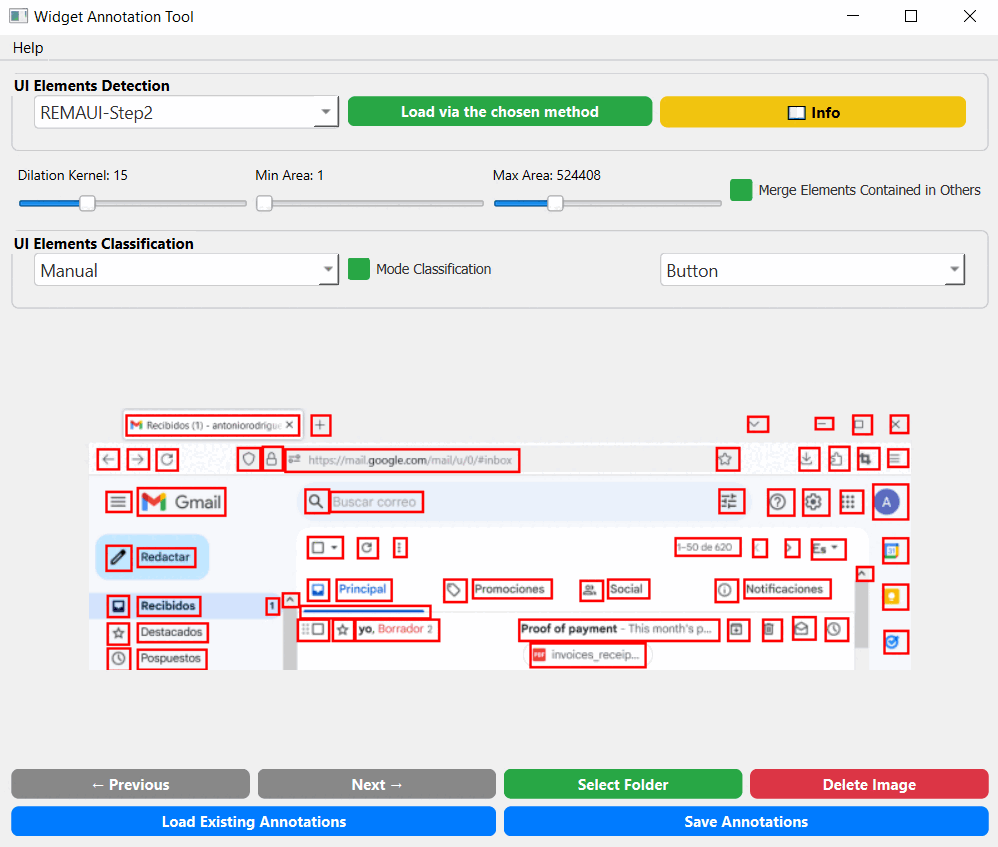}
\caption{Screenshot of the semi-automatic annotation tool \wat. The user selects a pre-labeling method (\textit{UI Elements Detection}), adjusts the parameters of the method with real-time feedback, and corrects or draws bounding boxes directly on the screenshot using interactive options.}
\Description{Screenshot of the semi-automatic annotation tool WidgetAnnotator. The user selects a pre-labeling method, adjusts the parameters, and corrects or draws bounding boxes directly on the screenshot using interactive options. Bounding boxes are represented as red rectangles.}
\label{fig:annotation_tool}
\end{figure}
\subsection{GUI element types\label{app:gui_element_types}}
List of GUI widgets:
\begin{enumerate}
\item The \textit{ToggleButton} class includes all binary state controls such as checkboxes, switches, and radio buttons.

\begin{figure}[H]
\centering
\includegraphics[width=0.9\columnwidth]{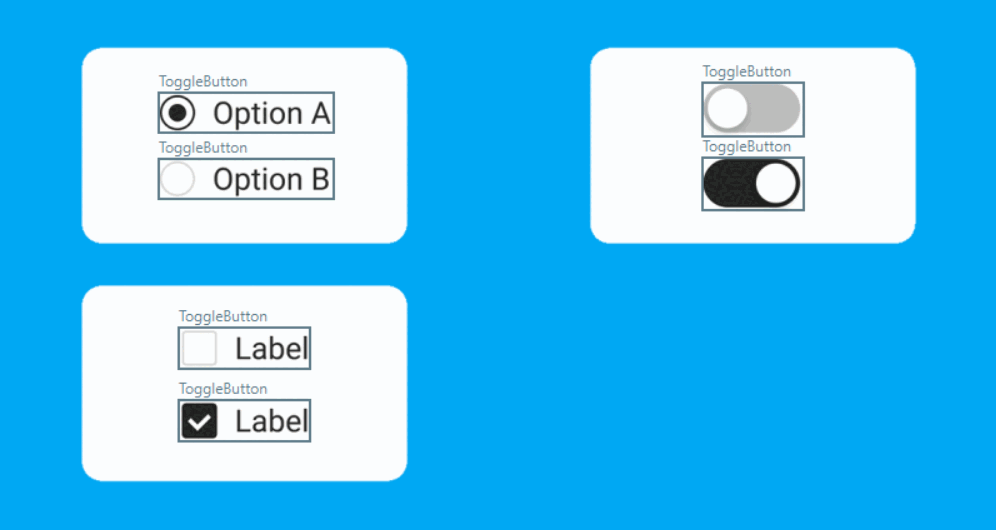}
\caption{Example of annotated ToggleButtons.}
\Description{Example of annotated ToggleButtons with gray bounding boxes around each button.}
\label{fig:togglebutton}
\end{figure}

\item The \textit{TextInput} class covers interactive fields that allow users to enter text when selected, such as form inputs or search bars.

\begin{figure}[H]
\centering
\includegraphics[width=0.9\columnwidth]{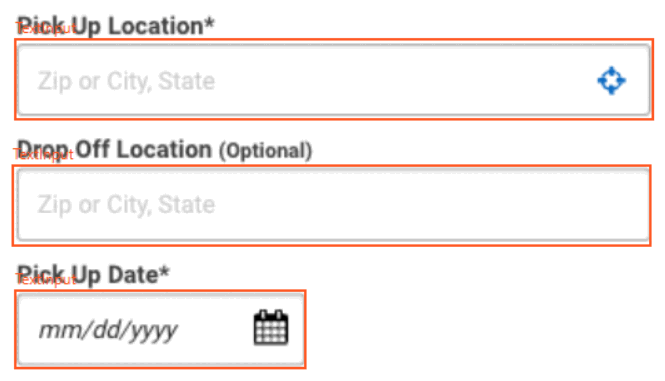}
\caption{Example of annotated TextInput fields.}
\Description{Example of annotated TextInput fields with red bounding boxes around each field.}
\label{fig:textinput}
\end{figure}

\item The \textit{Slider} class refers to sliding controls like volume adjusters or scrollbars; in such cases, we annotate the entire track along which the slider moves, excluding any side buttons, which are labeled as \textit{Button}.

\begin{figure}[H]
\centering
\includegraphics[width=0.9\columnwidth]{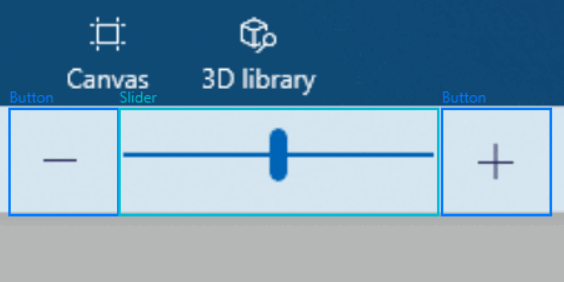}
\caption{Example of annotated slider.}
\Description{Example of annotated slider with blue bounding boxes around the slider and buttons.}
\label{fig:slider}
\end{figure}

\item  The \textit{Text} class covers regions containing textual content such as paragraphs, code blocks, or documentation. We do not annotate isolated short words unless they are visually emphasized (\eg large headings) or grouped together, to avoid confusion with single-word buttons.

\item The \textit{Hyperlink} class is restricted to clickable text contained within a \textit{Text} block, in order to distinguish it from single-word clickable buttons.

\begin{figure}[H]
\centering
\includegraphics[width=0.9\columnwidth]{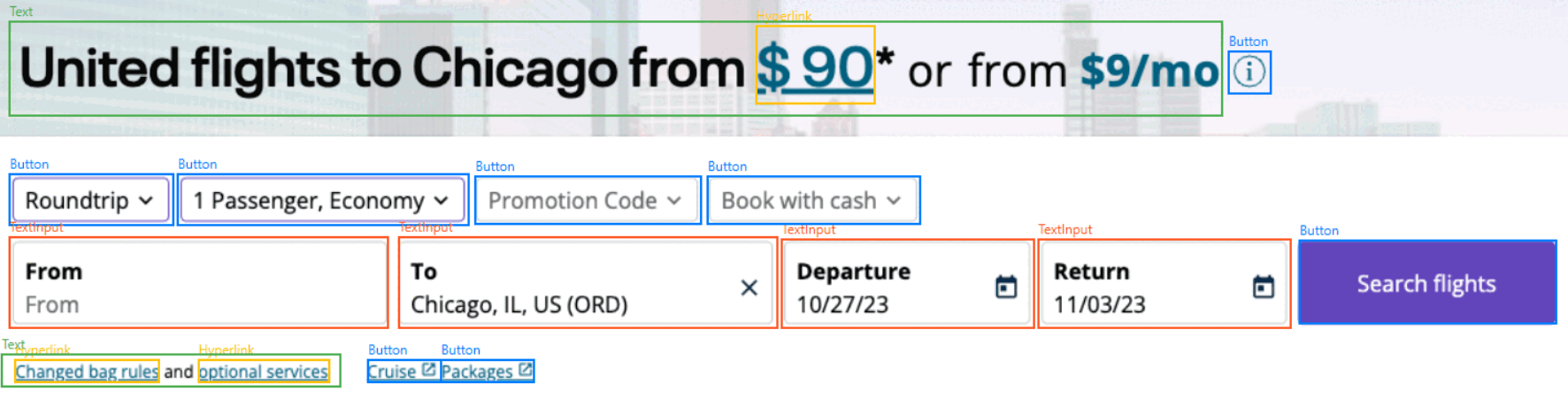}
\caption{Example of annotated hyperlinks and text blocks.}
\Description{Example of annotated hyperlinks and text blocks with orange bounding boxes around textInputs, blue around buttons and green around text blocks.}
\label{fig:hyperlink_text}
\end{figure}

\item The \textit{Button} class encompasses all other clickable components not included in the previous categories, such as dropdown triggers, tabs, or general-purpose buttons.

\begin{figure}[H]
\centering
\includegraphics[width=0.9\columnwidth]{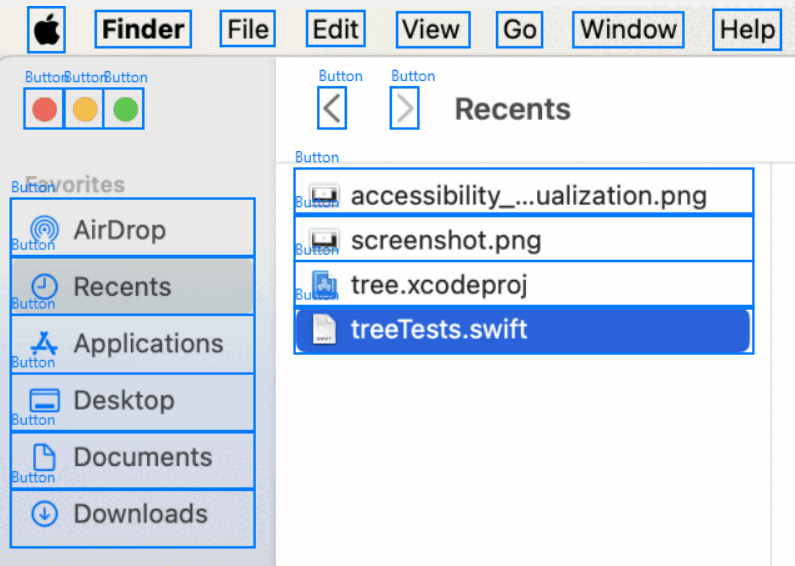}
\caption{Example of various annotated buttons.}
\Description{Example of various annotated buttons with blue bounding boxes around each button.}
\label{fig:button}
\end{figure}

\end{enumerate}

\subsection{Evaluation metrics for object detection\label{app:metrics}}
The evaluation metrics commonly used in the object detection literature are:

\begin{enumerate}
\item \textbf{Intersection over Union (IoU)}: This is the ratio between the number of pixels in the intersection of the ground truth object and the predicted object (true positive), and the number of pixels in their union. It measures the localization and size quality of the bounding boxes. See \autoref{fig:exemple_mobile}. The mean IoU (mIoU) represents the average IoU over all true positives.

\begin{figure}[H]
\centering
\includegraphics[width=0.5\columnwidth]{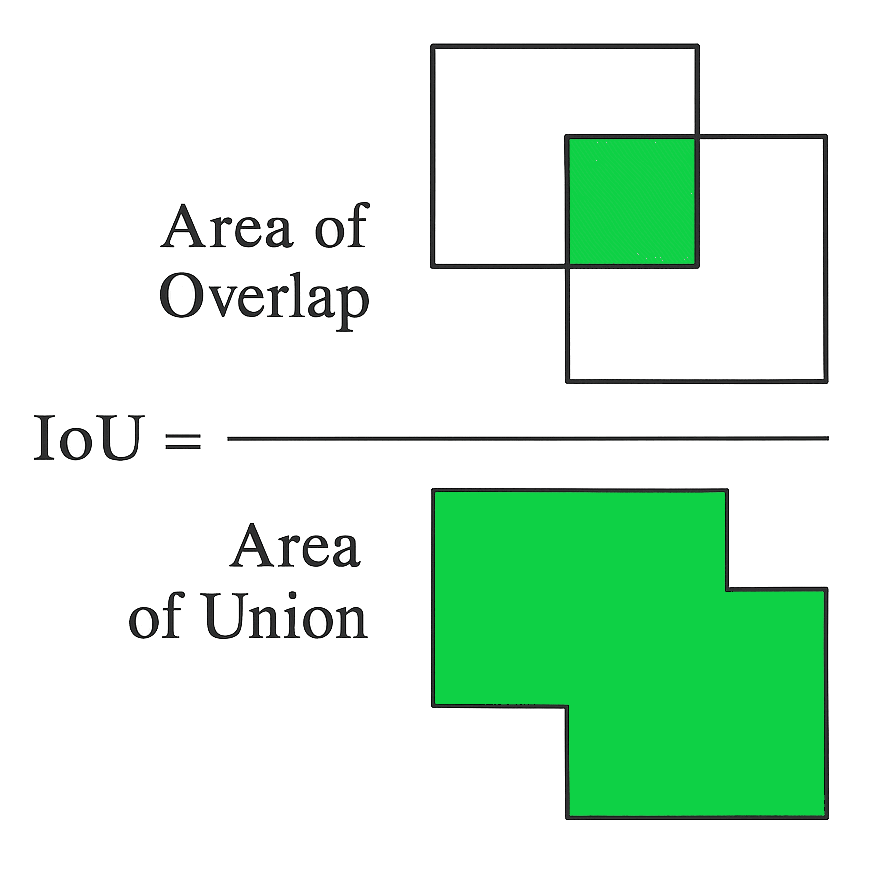}
\caption{Intersection over Union.}
\Description{Ilustration of Intersection over Union (IoU) with example bounding boxes.}
\label{fig:exemple_mobile}
\end{figure}

\item \textbf{Precision}: The number of correct detections (true positives) divided by the total number of detections. It reflects the model’s ability to avoid false positives.

\[
\text{Precision} = \frac{\text{True Positives}}{\text{True Positives} \text{ + } \text{False Positives}}
\]

\medskip

\item \textbf{Recall}: The number of correct detections (true positives) divided by the total number of ground truth instances (i.e., the sum of true positives and false negatives). It reflects the model’s ability to avoid missing targets.

\[
\text{Recall} = \frac{\text{True Positives}}{\text{True Positives} \text{ + } \text{False Negatives}}
\]

\medskip

\item \textbf{F1 Score}: The harmonic mean of precision and recall. It offers a balance between these two metrics and penalizes imbalances (if one is very low, the F1 score will also be low).

\[
\text{F1} = \frac{2 \cdot \text{Precision} \cdot \text{Recall}}{\text{Precision} \text{ + } \text{Recall}}
\]

\medskip

In practice, the F1 score can be plotted as a function of the confidence threshold to determine the score that best balances precision and recall.

\medskip

\item \textbf{Mean Average Precision mAP@0.5}: The average of the AP@0.5 across all classes, where AP@0.5 is the area under the precision-recall curve measured at IoU > 0.5. It provides an overall measure of model performance.

\medskip

\item \textbf{Mean Average Precision mAP@0.5:0.95}: Similar to mAP@0.5, but averaged over IoU thresholds ranging from 0.5 to 0.95 in increments of 0.05 (stricter). This gives better insight into how the model performs as the localization accuracy requirement increases.

\end{enumerate}

\subsection{Agreement with the primary annotator}
\label{app:primary_gt_agreement}

We randomly selected 15 images and asked three independent annotators to re-label them using the same guidelines and the same tool described in Section~\ref{subs:tool}. Their annotations were then compared to those of the primary annotator (the version used in the final dataset).

To quantify agreement, we use the F1-score and the mIoU, with matches defined using an IoU threshold of 0.5.

Table~\ref{tab:primarygt} reports these metrics for each of the 15 sampled images, for all three annotators, together with their means.

\begin{table}[H]
\centering
\caption{Agreement with the primary annotator for the three independent annotators.}
\Description{Table showing the agreement between the primary annotator and three independent annotators. For each of the 15 sampled images, the table reports the F1-score and the mean IoU for Ann1, Ann2, and Ann3, along with their average.}
\label{tab:primarygt}
\resizebox{\columnwidth}{!}{
\begin{tabular}{lcccccccc}
\toprule
\multirow{2}{*}{\textbf{Image ID}} &
\multicolumn{4}{c}{\textbf{F1}} &
\multicolumn{4}{c}{\textbf{mIoU}} \\
\cmidrule(lr){2-5} \cmidrule(lr){6-9}
& Ann1 & Ann2 & Ann3 & Mean & Ann1 & Ann2 & Ann3 & Mean \\
\midrule
img\_001 & 0.892 & 0.925 & 0.922 & 0.913 & 0.886 & 0.875 & 0.859 & 0.873 \\
img\_002 & 0.952 & 0.972 & 0.993 & 0.973 & 0.917 & 0.918 & 0.912 & 0.916 \\
img\_003 & 0.957 & 0.945 & 0.802 & 0.902 & 0.896 & 0.874 & 0.863 & 0.877 \\
img\_004 & 0.880 & 0.905 & 0.772 & 0.852 & 0.850 & 0.844 & 0.851 & 0.848 \\
img\_005 & 0.901 & 0.913 & 0.907 & 0.907 & 0.915 & 0.893 & 0.912 & 0.907 \\
img\_006 & 0.981 & 0.962 & 0.981 & 0.974 & 0.927 & 0.937 & 0.935 & 0.933 \\
img\_007 & 0.970 & 0.983 & 0.978 & 0.977 & 0.874 & 0.875 & 0.879 & 0.876 \\
img\_008 & 0.989 & 0.994 & 0.989 & 0.991 & 0.956 & 0.959 & 0.960 & 0.958 \\
img\_009 & 0.882 & 0.857 & 1.000 & 0.913 & 0.934 & 0.931 & 0.930 & 0.932 \\
img\_010 & 0.894 & 0.905 & 0.917 & 0.905 & 0.737 & 0.751 & 0.718 & 0.735 \\
img\_011 & 0.937 & 1.000 & 0.975 & 0.971 & 0.946 & 0.932 & 0.939 & 0.939 \\
img\_012 & 0.945 & 0.934 & 0.886 & 0.922 & 0.876 & 0.874 & 0.821 & 0.857 \\
img\_013 & 0.944 & 0.959 & 0.920 & 0.941 & 0.894 & 0.881 & 0.880 & 0.885 \\
img\_014 & 0.948 & 0.924 & 0.931 & 0.934 & 0.922 & 0.922 & 0.918 & 0.920 \\
img\_015 & 0.951 & 0.852 & 0.967 & 0.923 & 0.922 & 0.934 & 0.913 & 0.923 \\
\midrule
\textbf{Mean} & 0.935 & 0.935 & 0.929 & \textbf{0.933} & 0.897 & 0.893 & 0.886 & \textbf{0.892} \\
\bottomrule
\end{tabular}
}
\end{table}

\subsection{Full visual comparison}

A visual comparison of widget detection outputs from other methods and our trained YOLO26 model for three example screenshots is presented Figure \ref{fig:appendix_mosaic_comparison_labels}.

{ \setlength{\tabcolsep}{0pt}         \renewcommand{\arraystretch}{0}

\begin{figure*}\centering
\makebox[\textwidth][c]{
\resizebox{0.94\textwidth}{!}{\begin{tabular}{@{} >{\raggedright\arraybackslash}b{0.03\textwidth}
>{\centering\arraybackslash}p{0.35\textwidth}
>{\centering\arraybackslash}p{0.279\textwidth}
>{\centering\arraybackslash}p{0.313\textwidth} @{}}

& \textbf{Ubuntu example} & \textbf{macOs Example} & \textbf{Windows 10 Example} \\[1pt]

\raisebox{2ex}{\rotatebox{90}{\textbf{Ground truth}}} & \includegraphics[width=\linewidth]{supplementary_materials/GUI_Detection_Examples_YOLO_and_OtherMethods/figures/GT_ex1.png}
& \includegraphics[width=\linewidth]{supplementary_materials/GUI_Detection_Examples_YOLO_and_OtherMethods/figures/GT_ex3.png}
& \includegraphics[width=\linewidth]{supplementary_materials/GUI_Detection_Examples_YOLO_and_OtherMethods/figures/GT_ex2.png} \\[0pt]

\raisebox{2ex}{\rotatebox{90}{\textbf{Our YOLO26}}} & \includegraphics[width=\linewidth]{supplementary_materials/GUI_Detection_Examples_YOLO_and_OtherMethods/figures/OurModel_ex1.png}
& \includegraphics[width=\linewidth]{supplementary_materials/GUI_Detection_Examples_YOLO_and_OtherMethods/figures/OurModel_ex3.png}
& \includegraphics[width=\linewidth]{supplementary_materials/GUI_Detection_Examples_YOLO_and_OtherMethods/figures/OurModel_ex2.png} \\[0pt]

\rotatebox{90}{\textbf{OmniParser YOLO}} & \includegraphics[width=\linewidth]{supplementary_materials/GUI_Detection_Examples_YOLO_and_OtherMethods/figures/OmniParser_ex1.png}
& \includegraphics[width=\linewidth]{supplementary_materials/GUI_Detection_Examples_YOLO_and_OtherMethods/figures/OmniParser_ex3.png}
& \includegraphics[width=\linewidth]{supplementary_materials/GUI_Detection_Examples_YOLO_and_OtherMethods/figures/OmniParser_ex2.png} \\[0pt]

\raisebox{4ex}{\rotatebox{90}{\textbf{REMAUI}}} & \includegraphics[width=\linewidth]{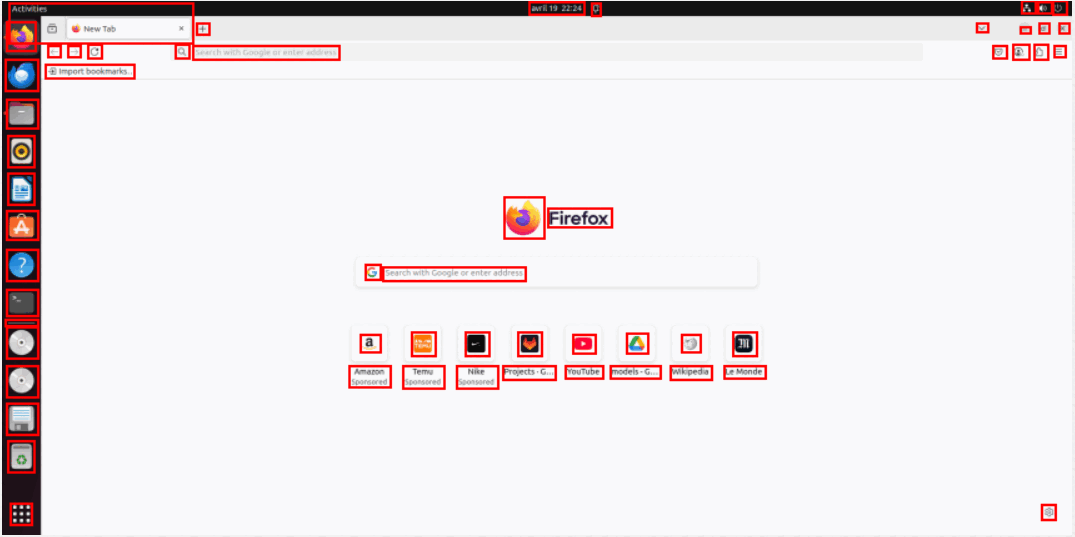}
& \includegraphics[width=\linewidth]{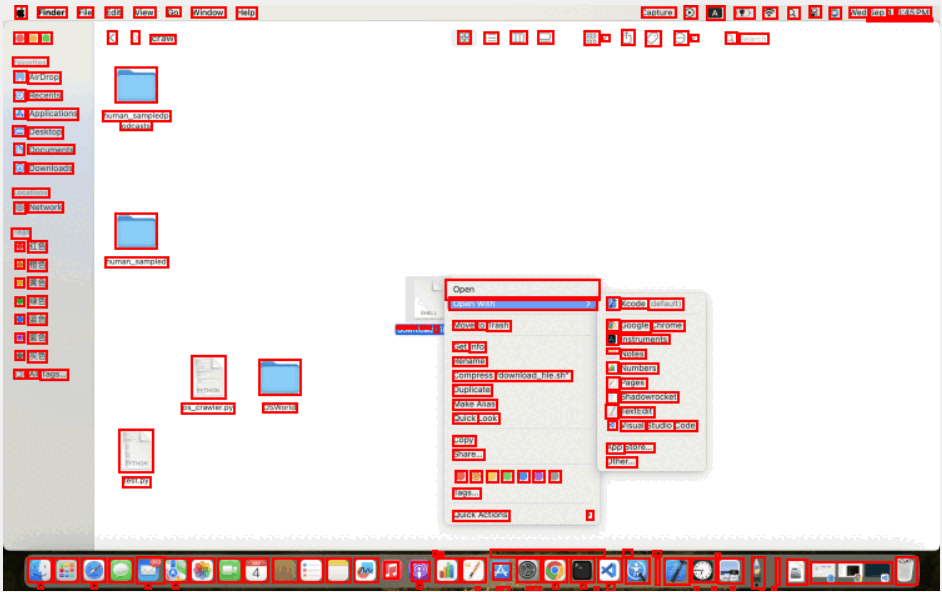}
& \includegraphics[width=\linewidth]{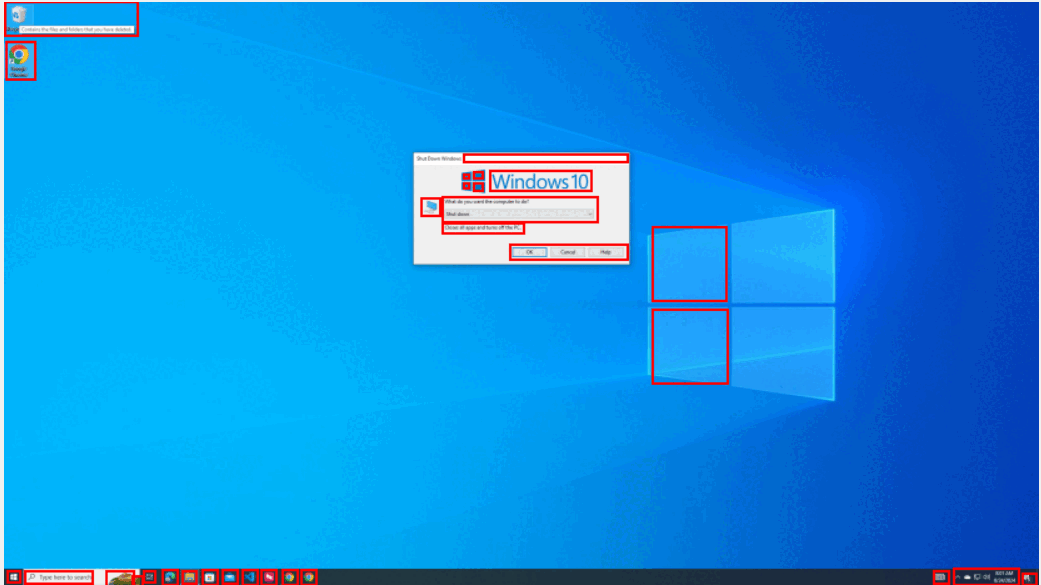} \\[0pt]

\raisebox{5ex}{\rotatebox{90}{\textbf{UIED}}} & \includegraphics[width=\linewidth]{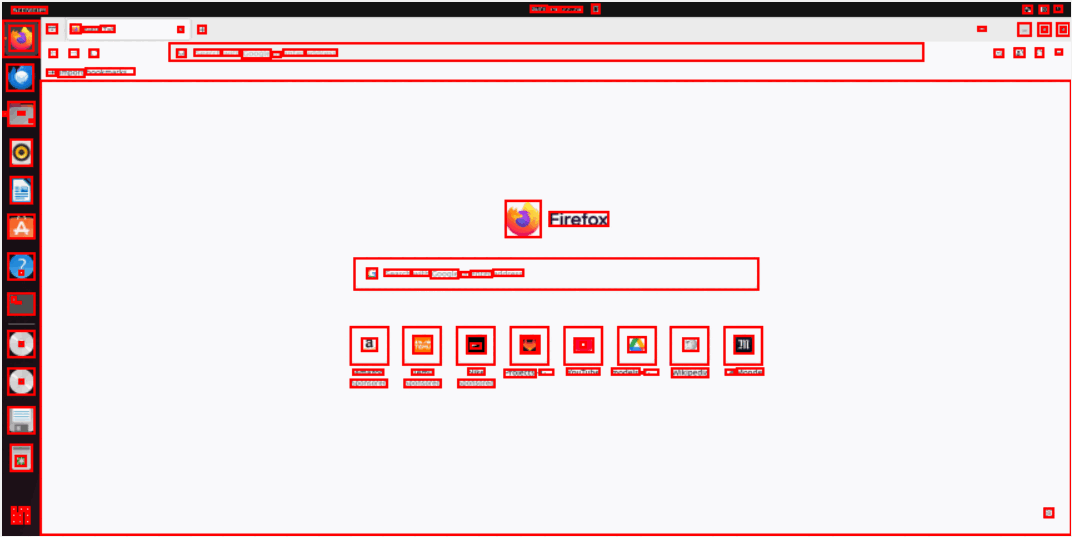}
& \includegraphics[width=\linewidth]{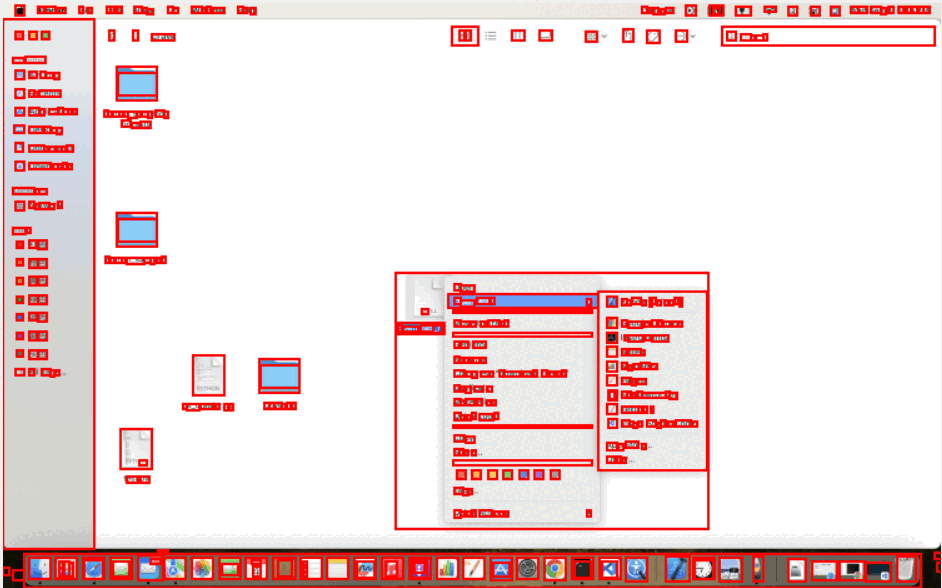}
& \includegraphics[width=\linewidth]{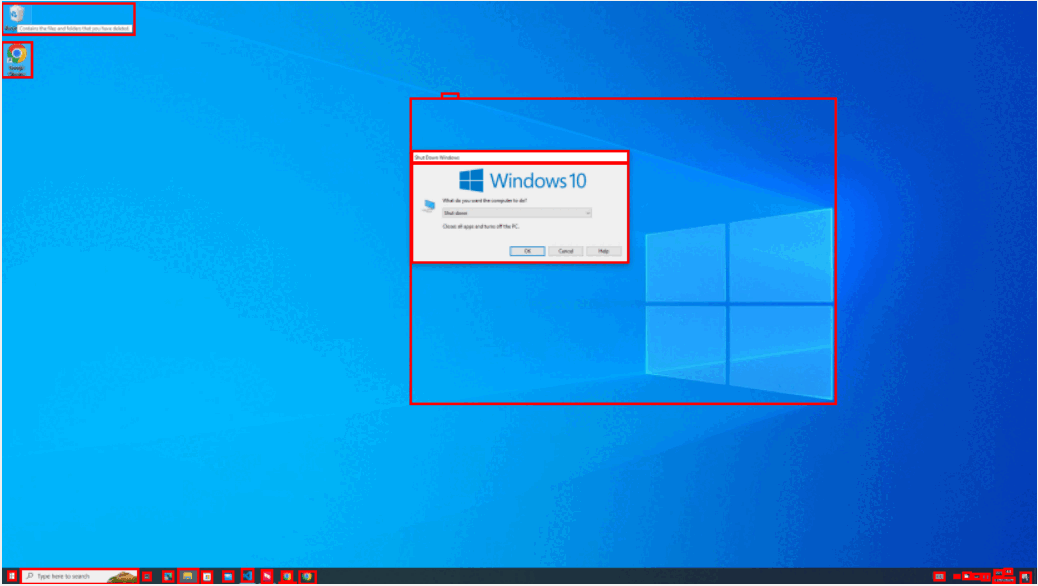} \\[0pt]

\raisebox{3ex}{\rotatebox{90}{\textbf{MobileSAM}}} & \includegraphics[width=\linewidth]{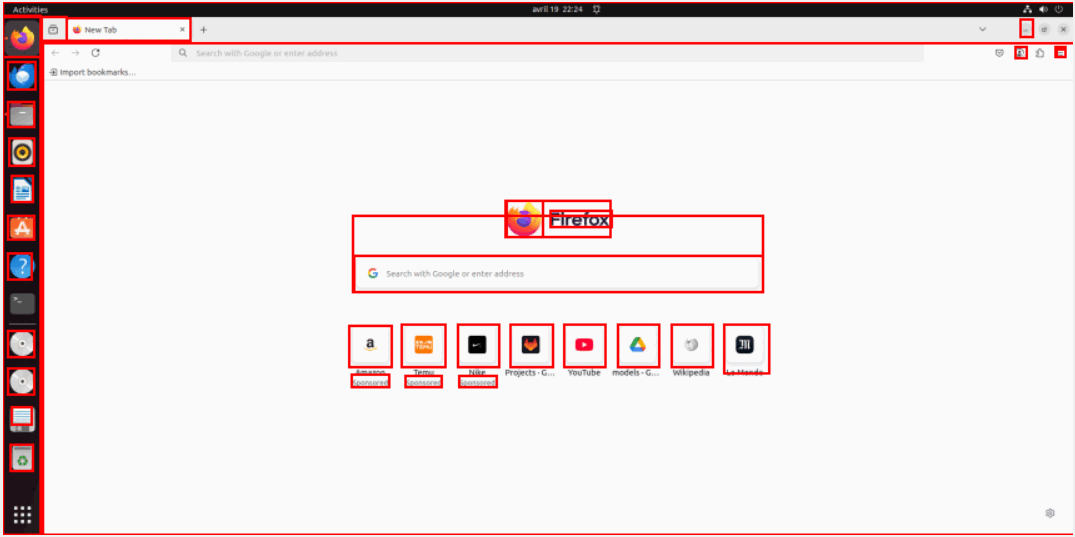}
& \includegraphics[width=\linewidth]{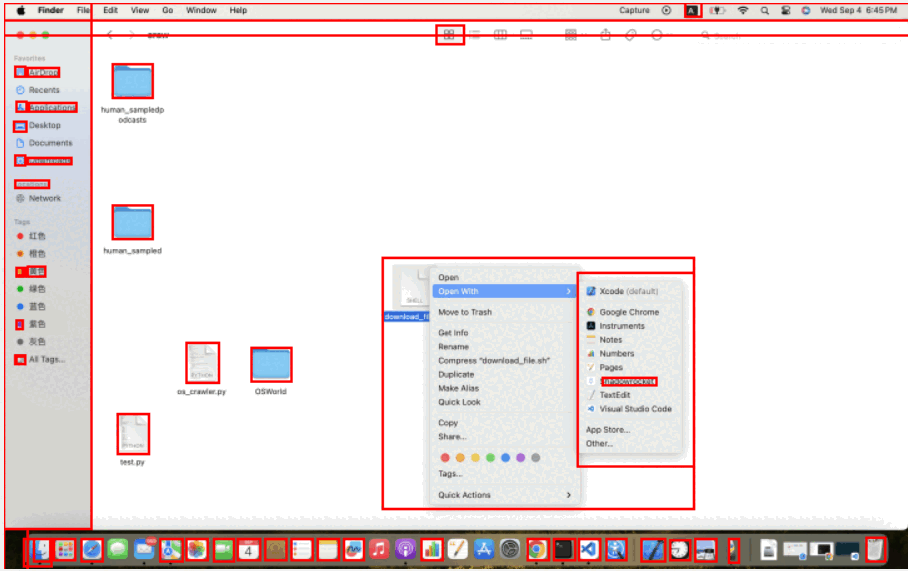}
& \includegraphics[width=\linewidth]{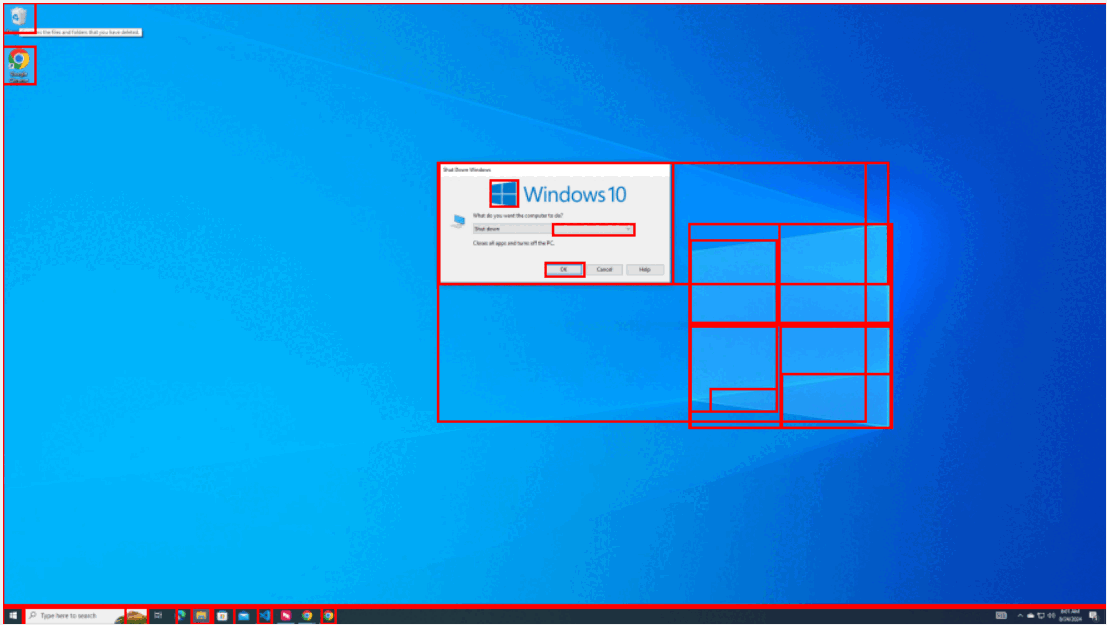} \\[0pt]

\raisebox{1ex}{\rotatebox{90}{\textbf{Martinez YOLO}}} & \includegraphics[width=\linewidth]{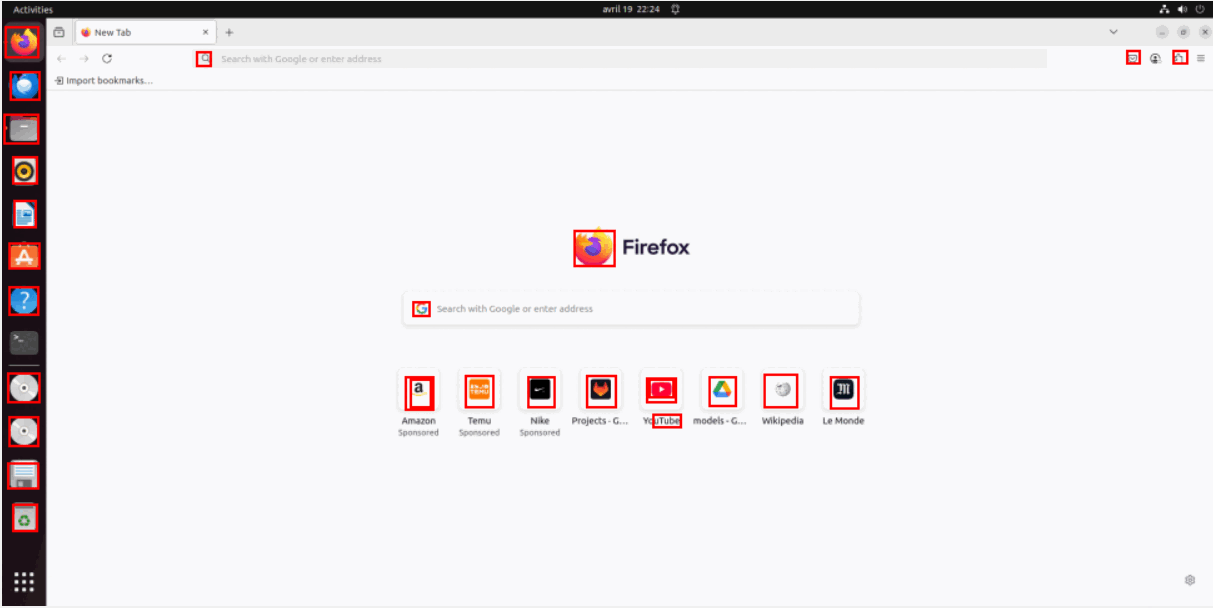}
& \includegraphics[width=\linewidth]{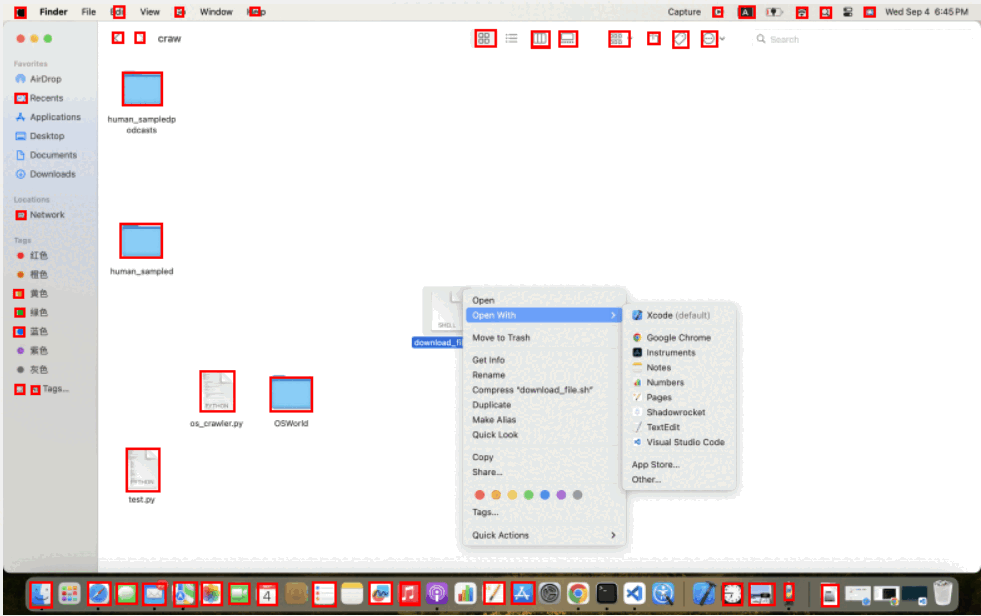}
& \includegraphics[width=\linewidth]{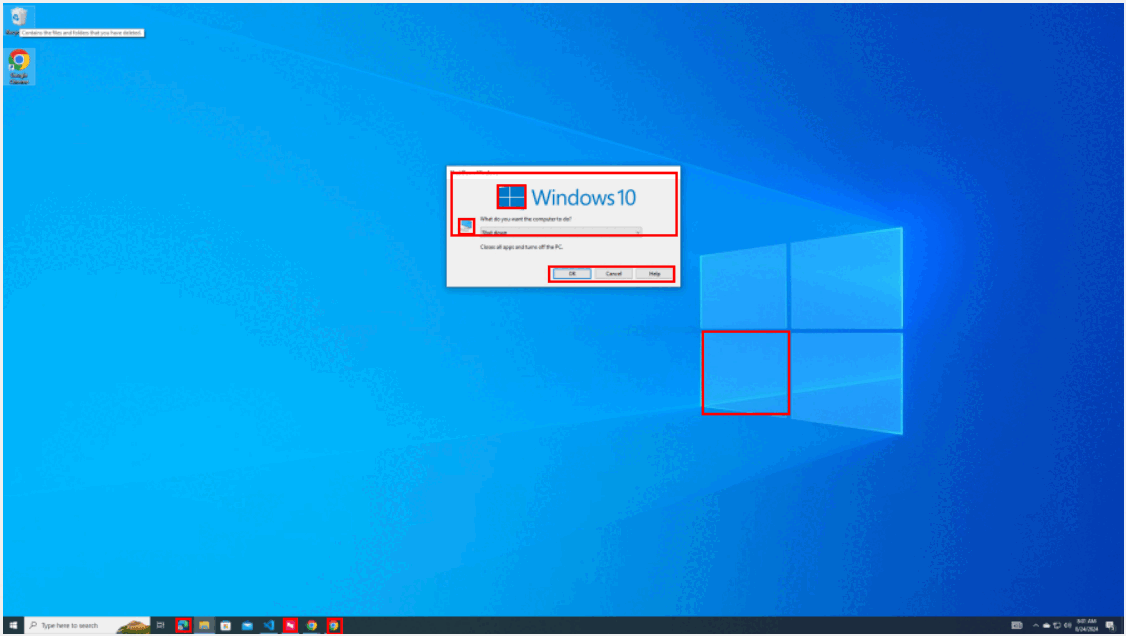} \\

\end{tabular}
}
}
\caption{Visual comparison of widget detection outputs from other methods and our trained YOLO26 model for three example screenshots: a Firefox browser window in Ubuntu (left), a macOs desktop with white background (middle) and a Windows desktop with blue background and a screensaver that can lead to false detections (right).}
\Description{Visual comparison of widget detection outputs from other methods and our trained YOLO26 model for three example screenshots: a Firefox browser window in Ubuntu (left), a macOs desktop with white background (middle) and a Windows desktop with blue background and a screensaver that can lead to false detections (right). The first row shows the ground truth annotations, followed by rows showing the outputs from our YOLO26 model, OmniParser YOLO, REMAUI, UIED, MobileSAM, and Martinez-Rojas YOLO respectively. Each method's output is displayed in a separate row for each of the three examples.}
\label{fig:appendix_mosaic_comparison_labels}
\end{figure*}

}

\subsection{Other performances}\label{other_perfs}

\begin{table}[H]
\centering
\caption{Performance of OmniParser (YOLO11m-1280) on progressively smaller UI elements (filtered by their YOLO-pixel size).}
\Description{Table showing the performance of OmniParser (YOLO11m-1280) on UI elements grouped by their YOLO-pixel size. The rows correspond to all widgets, and to widgets below the 50th, 30th, and 10th percentiles. For each subset, the table reports Precision, Recall, F1-score, mAP@0.5, mAP@0.5:0.95, and mIoU.}
\label{tab:small_elements_omni}
\resizebox{\columnwidth}{!}{
\begin{tabular}{lcccccc}
\toprule
\textbf{Size subset} & \textbf{Precision} & \textbf{Recall} & \textbf{F1} & \textbf{mAP@0.5} & \textbf{mAP@0.5:0.95} & \textbf{mIoU} \\
\midrule
All sizes      & 0.753 & 0.651 & 0.698 & 0.693 & 0.391 & 0.787 \\
Percentile 50  & 0.588 & 0.478 & 0.527 & 0.383 & 0.185 & 0.757 \\
Percentile 30  & 0.392 & 0.390 & 0.391 & 0.206 & 0.087 & 0.742 \\
Percentile 10  & 0.205 & 0.214 & 0.209 & 0.056 & 0.027 & 0.759 \\
\bottomrule
\end{tabular}
}
\end{table}

\begin{table}[H]
\centering
\caption{Performance of REMAUI on progressively smaller UI elements (filtered by their YOLO-pixel size).}
\Description{Table showing the performance of REMAUI on UI elements grouped by their YOLO-pixel size. he rows correspond to all widgets, and to widgets below the 50th, 30th, and 10th percentiles. For each subset, the table reports Precision, Recall, F1-score, mAP@0.5, mAP@0.5:0.95, and mIoU.}
\label{tab:small_elements_remaui}
\resizebox{\columnwidth}{!}{
\begin{tabular}{lcccccc}
\toprule
\textbf{Size subset} & \textbf{Precision} & \textbf{Recall} & \textbf{F1} & \textbf{mAP@0.5} & \textbf{mAP@0.5:0.95} & \textbf{mIoU} \\
\midrule
All sizes      & 0.430 & 0.484 & 0.455 & N/A & N/A & 0.749 \\
Percentile 50  & 0.333 & 0.427 & 0.374 & N/A & N/A & 0.742 \\
Percentile 30  & 0.227 & 0.388 & 0.287 & N/A & N/A & 0.751 \\
Percentile 10  & 0.160 & 0.256 & 0.197 & N/A & N/A & 0.801 \\
\bottomrule
\end{tabular}
}
\end{table}

\begin{table}[H]
\centering
\caption{Performance of \technique (YOLO26n-640) on progressively smaller UI elements (filtered by their YOLO-pixel size).}
\Description{Table showing the performance of TargetFinder trained with a 640×640 input resolution on UI elements grouped by their YOLO-pixel size. The rows correspond to all widgets, and to widgets below the 50th, 30th, and 10th percentiles. For each subset, the table reports Precision, Recall, F1-score, mAP@0.5, mAP@0.5:0.95, and mIoU.}
\label{tab:small_elements_640}
\resizebox{\columnwidth}{!}{
\begin{tabular}{lcccccc}
\toprule
\textbf{Size subset} & \textbf{Precision} & \textbf{Recall} & \textbf{F1} & \textbf{mAP@0.5} & \textbf{mAP@0.5:0.95} & \textbf{mIoU} \\
\midrule
All sizes      & 0.936 & 0.840 & 0.885 & 0.899 & 0.697 & 0.872 \\
Percentile 50  & 0.901 & 0.700 & 0.788 & 0.674 & 0.470 & 0.839 \\
Percentile 30  & 0.805 & 0.603 & 0.689 & 0.526 & 0.344 & 0.818 \\
Percentile 10  & 0.691 & 0.303 & 0.421 & 0.252 & 0.133 & 0.764 \\
\bottomrule
\end{tabular}
}
\end{table}

\appendix

\end{document}